\documentclass[aps,pra,twocolumn,a4paper,showpacs,nofootinbib]{revtex4}
\usepackage{pifont}
\usepackage{amsmath}
\usepackage{amssymb}
\usepackage{amsfonts}
\usepackage{times,txfonts}
\usepackage{bbold}
\usepackage{color}
\usepackage{bm}
\usepackage{graphicx}
\usepackage{epsfig}
\usepackage{verbatim}
\usepackage{epstopdf}

\definecolor{orange}{rgb}{1,0.5,0}

\begin{document}

\title{Fractional resonances in the atom-optical $\delta$-kicked accelerator}

\author{M. Saunders}
\affiliation{Department of Physics, Durham University, Rochester Building, South Road, Durham DH1 3LE, United Kingdom}

\author{P. L. Halkyard}
\affiliation{Department of Physics, Durham University, Rochester Building, South Road, Durham DH1 3LE, United Kingdom}

\author{K. J. Challis}
\affiliation{Lundbeck Foundation Theoretical Center for Quantum System Research, Department of Physics and Astronomy, University of Aarhus, DK-8000 \AA rhus  C, Denmark}

\author{S. A. Gardiner}
\affiliation{Department of Physics, Durham University, Rochester Building, South Road, Durham DH1 3LE, United Kingdom}

\date{\today}

\begin{abstract}
We consider resonant dynamics in a dilute atomic gas falling under gravity through a periodically pulsed standing-wave laser field.  Our numerical calculations are based on a Monte Carlo method for an incoherent mixture of noninteracting plane waves, and show that quantum resonances are highly sensitive to the relative acceleration between the atomic gas and the pulsed optical standing wave.  For particular values of the atomic acceleration, we observe fractional resonances.  We investigate the effect of the initial atomic momentum width on the fractional resonances, and quantify the sensitivity of fractional resonances to thermal effects.
\end{abstract}


\maketitle

\section{Introduction}

The atom-optical kicked rotor can be realized by subjecting a dilute atomic gas to a pulsed optical standing wave.  This system has become well established as a convenient way to study quantum chaotic dynamics \cite{Oskay2000,Sadgrove2004,Bharucha1999,dArcy2001a,Oberthaler1999,Godun2000,Schlunk2003a,Schlunk2003b,Ma2004,Buchleitner2006,Behinaein2006,Guarneri2008,Moore1995,Ammann1998a,dArcy2001b,Moore1994,Klappauf1999,Steck2000,Milner2000,Oskay2003,Vant2000,Doherty2000,Kanem2007, dArcy2003,Duffy2004b,Ryu2006,Szriftgiser2002,Ammann1998b,Vant1999,Williams2004,Duffy2004a,Tonyushkin2008, Wu2008,Jones2004}.  In particular, quantum resonances and antiresonances \cite{Casati1979,Izrailev1979,Izrailev1980,Dana1995,Dana1996,Dana2005,Dana2006a,Dana2006b,Wimberger2003,Wimberger2004,Wimberger2005a} are a dramatic signature of quantum chaos \cite{Reichl2004,Haake2001,Gutzwiller1990}, and these phenomena have been observed and studied in detail in cold atom kicked-rotor experiments \cite{Bharucha1999,Oskay2000, dArcy2001a,Sadgrove2004}.  One of the features of such experiments is that the optical standing wave is often oriented vertically, allowing the atom cloud to fall under gravity during the optical pulses.  This realizes the quantum kicked accelerator  \cite{dArcy2001a}, and has led to the observation of quantum accelerator modes \cite{Oberthaler1999,Godun2000,dArcy2001a,Schlunk2003a,Schlunk2003b,Ma2004,Buchleitner2006,Behinaein2006,Guarneri2008} which are closely related to quantum resonances.  However, quantum resonances themselves have not been specifically investigated for the quantum delta-kicked accelerator.

In this paper, we consider the atom-optical quantum kicked accelerator where the linear time-independent potential, typically provided by the gravitational acceleration, can be freely chosen.  This is possible in experiments with either vertical or horizontal orientation of the optical standing wave, as an appropriate variation in the phase of the optical standing wave can induce an effective gravitational acceleration.  We find that the relative acceleration between the atom cloud and the pulsed optical standing wave can lead to {\it fractional resonances}.  We characterize these fractional resonances and consider in detail the effect of the initial atomic momentum width on the fractional resonant dynamics.  This generalizes our previous discussion on the effect of temperature on quantum resonances and antiresonances in the atom-optical quantum kicked rotor \cite{Saunders2007}.

This paper is organized as follows.  In Sec.\ \ref{sec:atomic_system}, we derive the quantum $\delta$-kicked accelerator Hamiltonian for the system of a two-level atom falling through an optical standing wave.  We then consider the time evolution of the system based on a Floquet operator approach.  Our treatment closely follows the derivation presented in Sec. II of Saunders \textit{et al}.\ \cite{Saunders2007}. In Sec.\ \ref{sec:fractional} we consider fraction resonances in the zero-temperature limit and for the case of a broad, thermal atomic momentum distribution.  As in our previous work \cite{Saunders2007}, we use the evolution of the momentum moments to characterize the atomic dynamics.  In Sec.\ \ref{sec:temperature} we consider the dynamics of a plane wave state and investigate the effect of the initial momentum on the system behavior.  From that discussion we develop an understanding of the temperature dependence of fractional resonances.  We conclude  in Sec.\ \ref{sec:conclusions}.

\section{Atomic System \label{sec:atomic_system}}

\subsection{Model Hamiltonian}

\subsubsection{Freely-falling two-level atom in a laser field}

We consider a cloud of trapped and laser-cooled thermal alkali atoms.  The cloud is released from all external fields, and then falls through a vertically aligned pulsed sinusoidal potential \cite{dArcy2001a,Oberthaler1999,Godun2000,Schlunk2003a,Schlunk2003b,Ma2004,Buchleitner2006,Behinaein2006} formed by two counter-propagating laser beams \cite{Denschlag2002}.  We neglect interatomic collisions, as is reasonable for a dilute thermal gas.  For the case of an initially Bose-condensed sample where atom-atom interactions can be significant, it is possible to tune the scattering length to zero \cite{Gustavsson2007} near a Feshbach resonance \cite{Inouye1998,Roberts1998,Kohler2006,Kraemer2004}.

We describe the system using a one-dimensional model along the vertical axis.  As in \cite{Saunders2007}, the Hamiltonian $\hat{H}$ describes a single two-level atom of mass $M$, with internal ground state $|g\rangle$ and excited state $|e\rangle$, but now the centre-of-mass motion is influenced by a linear gravitational potential.  The internal atomic levels are coupled by two equal-frequency laser fields with phases $\phi_{1}$ and $\phi_{2}$, respectively, and laser wave-vector magnitude in the vertical direction $k_L$. We consider the lasers to be of equal intensity, and far detuned so that spontaneous emission can be neglected.  As in \cite{Saunders2007}, we work in a rotating frame defined by $\hat{U}_1 = \exp(i[\omega_{L}|e\rangle\langle e| -\omega_{0}(|e\rangle \langle e| + |g\rangle\langle g|)/2]t)$, and so
\begin{equation}
\begin{split}
\hat{H} = & \hbar\Delta |e\rangle \langle e| + \frac{\hat{p}^{2}}{2M} +
Mg \hat{z} \\
& + \frac{1}{2} \hbar \Omega_{R}\cos(k_L \hat{z}+\phi_{r}/2) \left[ e^{i\phi_{s}}|e\rangle \langle g | + \mbox{H.c.} \right].
\end{split}
\label{Hprime}
\end{equation}
Here $\hat{z}$ is the vertical centre-of-mass atomic position operator with $\hat{p}$ its conjugate momentum, $t$ is the time,  $g$ is the local gravitational acceleration, $\phi_{r}=\phi_{1}-\phi_{2}$, $\phi_{s}=(\phi_{1}+\phi_{2})/2$, and  the detuning $\Delta = \omega_{0}-\omega_{L}$, with $\hbar\omega_{0}$ the energy difference between the internal atomic levels and $\omega_{L}$ the laser frequency.

\subsubsection{Transformation to an accelerating frame \label{sec:walking_wave}}

In contrast to \cite{Saunders2007}, we consider the laser phases $\phi_1$ and $\phi_2$ to be time-dependent.  In particular, we choose $\phi_{r}= k_L a_{\phi}t^{2}$ so that the laser field forms a ``walking wave'' with acceleration $-a_{\phi}$ \cite{Denschlag2002}.  This allows for the possibility of exactly negating the effect of the gravitational force by taking $a_{\phi}=g$, and for the versatility of investigating different effective values of the gravitational acceleration by choosing $a_{\phi}$ at our convenience.  The relative phase $\phi_{r}$ can be tuned with an accuracy on the order of one part in a million \cite{Denschlag2002}, allowing for precise control over the effective gravitational acceleration \cite{Ma2004}. 

The Hamiltonian (\ref{Hprime}) can be transformed to the frame accelerating with the walking wave by implementing the unitary transformation $\hat{U}_2=\exp(i[Ma_{\phi}\hat{z}t -a_{\phi}\hat{p}t^{2}/2+\xi(t)]/\hbar)$, where for convenience $\xi(t)=Ma_{\phi}t^{3}(a_{\phi}-2g)/12$.  This yields
\begin{equation}
\begin{split}
\hat{H}' = & \hbar\Delta |e\rangle \langle e| +\frac{ \hat{p}^{2}}{2M} + Ma\hat{z} \\
& + \frac{1}{2} \hbar \Omega_{R}\cos(k_L \hat{z})\left[ e^{i\phi_{s}}|e\rangle \langle g | + \mbox{H.c.} \right],
\end{split}
\label{H_pp}
\end{equation}
where we have used $\hat{U}_2\hat{z}\hat{U}_2^{\dagger} = \hat{z} - a_{\phi}t^{2}/2 $ and $ \hat{U}_2\hat{p}\hat{U}_2^{\dagger} = \hat{p} - Ma_{\phi}t $, 
and we have defined $a \equiv g-a_{\phi}$.  It is evident from Eq.\ (\ref{H_pp}) that, in the frame defined by $\hat{U}_{2}$, $a$ assumes the role of an effective gravitational acceleration.

In the far-detuned limit ($\Omega_{R}/\Delta \ll 1$), the excited state~$|e\rangle$ can be adiabatically eliminated~\cite{Saunders2007,Meystre1991}.  By also invoking the unitary transformation $\hat{U}_3=\exp(-i\Omega_R^2|g\rangle \langle g |t/8\Delta)$, the Hamiltonian simplifies to 
\begin{equation}
\hat{H}'' =\frac{ \hat{p}^{2}}{2M} + Ma\hat{z} - \frac{\hbar \Omega_{R}^{2}}{8\Delta}\cos(K\hat{z}),
\label{Hp_final}
\end{equation}
where $K=2k_L$, and the atoms are assumed always to be in the internal ground state.  In an exactly analogous fashion to \cite{Saunders2007}, we consider the laser standing-wave amplitude to be periodically pulsed, with period $T$ and pulse duration $t_{p}$.  For sufficiently short pulses \cite{Moore1995,Klappauf1999,Saunders2007}, It is then possible to approximate the Hamiltonian by
\begin{equation}
\hat{H}_{\rm \delta ka} =\frac{ \hat{p}^{2}}{2M} + Ma\hat{z} - \hbar
\phi_d \cos(K\hat{z})\sum_{n=0}^{\infty}\delta (t-nT),
\label{H_final}
\end{equation}
where we have defined an effective potential depth $\phi_{d} \equiv \Omega_{R}^{2}t_{p}/8\Delta$.  A system governed by Hamiltonian~(\ref{H_final}) is known as a $\delta$-kicked accelerator \cite{dArcy2001a}.

\subsubsection{Transformation to a spatially-periodic Hamiltonian \label{sec:cons_quasimom}}

It is convenient to transform Hamiltonian (\ref{H_final}) to a frame in which it takes a spatially periodic form so that we can apply Bloch theory \cite{Kittel1996, Bach2005, Fishman2002, Fishman2003, Saunders2007}.  This is accomplished by a unitary transformation defined by the operator \cite{Fishman2002,Fishman2003,Bach2005}
\begin{equation}
\hat{U}_4(t) = \exp(iMa\hat{z}t/\hbar).
\label{U_spa_per}
\end{equation}
The transformed Hamiltonian is then
\begin{equation}
\tilde{H}_{\rm \delta k a}=\frac{(\hat{p}-Mat)^{2}}{2M} - \hbar\phi_{d}
\cos(K\hat{z})\sum_{n=0}^{\infty}\delta(t-nT).
\label{Hamiltonian_gauge}
\end{equation}
We expand the position and momentum operators into discrete and continuous components, i.e.,
\begin{equation}
\hat{z} =   K^{-1}(2\pi \hat{l} +\hat{\theta}), 
\quad
\hat{p}  =  \hbar K(\hat{k} + \hat{\beta}),
\label{spatial_periodicity}
\end{equation}
where the eigenvalues of $\hat{l}$ and $\hat{k}$ are integers, the eigenvalues of $\hat{\theta}\in [-\pi,\pi)$ and the eigenvalues of the quasimomentum operator $\hat{\beta} \in \left[-1/2, 1/2\right)$.  Substituting the reformulated operators into Hamiltonian~(\ref{Hamiltonian_gauge}) yields
\begin{equation}
 \tilde{H}_{\rm \delta ka }= \frac{[\hbar K(\hat{k}+\hat{\beta})-Mat]^{2}}{2M}
- \hbar\phi_{d}
\cos ( \hat{\theta} ) \sum_{n=0}^{\infty}\delta(t-nT).
\label{Hamiltonian_k_beta}
\end{equation}

The quasimomentum operator $\hat{\beta}$ commutes with Hamiltonian~(\ref{Hamiltonian_k_beta}) \cite{Bach2005, Saunders2007}.   Therefore, the quasimomentum is conserved in the \textit{accelerating frame}, and the laser field induces coupling only between momentum eigenstates differing in momenta by integer multiples of $\hbar K$, i.e., integer multiples of two photon recoils \cite{dArcy2001a}.  

\subsection{Time evolution \label{sec:time_evolution}}

\subsubsection{Transformed Floquet operator in the accelerating frame}

For a time-periodic system, the unitary time-evolution operator for a single temporal period is known as the Floquet operator.  Hamiltonian~(\ref{H_final}) has temporal period $T$ and we choose to define the Floquet operator from just before one kick to just before the next, i.e., 
\begin{equation}
\hat{F}=\exp \left(-\frac{i}{\hbar} \left[
\frac{\hat{p}^{2}}{2M} + Ma\hat{z} \right]T \right) \exp \left(
i\phi_{d} \cos(K\hat{z}) \right).
\label{eq:Floquet}
\end{equation}

In the accelerating frame defined by Eq.\ (\ref{U_spa_per}), the Hamiltonian~(\ref{Hamiltonian_gauge}) is not periodic in time, so strictly speaking there is no Floquet operator.  However, it is useful to consider the \textit{transformed} Floquet operator which is the time-evolution operator describing kick-to-kick dynamics of the system in the accelerating frame.  The Floquet operator~(\ref{eq:Floquet}), transformed according to the unitary operator~(\ref{U_spa_per}), becomes
\begin{equation}
\tilde{F}(nT,[n-1]T)=\hat{U}_4(nT)\hat{F}\hat{U}_4^\dagger([n-1]T).
\label{transform_floquet}
\end{equation}
The time variables on the left-hand side are required to indicate explicitly that the transformed Floquet operator corresponds to the system evolution from just before the $(n-1)$th kick to just before the $n$th kick \cite{Bach2005}.

To evaluate the transformed Floquet operator (\ref{transform_floquet}), 
it is convenient to separate the terms 
that depend on $\hat{z}$ from those that depend on $\hat{p}$.  The
Floquet operator~(\ref{eq:Floquet}) can be rewritten as
\begin{equation}
\begin{split}
 \hat{F}  = & \exp\left(-\frac{iMa^2T^3}{6\hbar}\right) \exp\left(-\frac{i}{\hbar}\left[\frac{\hat{p}^2}{2M}T+\frac{\hat{p}a}{2}T^{2}\right]\right) \\
&\times\exp\left(-\frac{iMa\hat{z}T}{\hbar} + i\phi_{d} \cos(K\hat{z}) \right),
\label{first_floquet}
\end{split}
\end{equation}
as shown in Appendix~\ref{app:factor}.  Substituting Eq.\ (\ref{first_floquet}) into Eq.\ (\ref{transform_floquet}), then using 
$\hat{U}_{4}(nT)\hat{p}\hat{U}_{4}^{\dagger}(nT)=\hat{p}-ManT$
and simplifying, we find that
\begin{equation}
\begin{split}
 \tilde{F}(nT,[n-1]T) = & \exp\left( -\frac{iMa^{2}}{6\hbar}\left[3n^{2}-3n+1
\right]T^{3} \right) \\
& \times \exp\left( -\frac{i}{\hbar} \left[ \frac{\hat{p}^{2}}{2M}T-\frac{\hat{p}a}{2}(2n-1)T^{2} \right] \right) \\
& \times \exp\left( i\phi_{d} \cos(K\hat{z}) \right).
\label{Ftilde}
\end{split}
\end{equation}
The first exponential term in Eq.\ (\ref{Ftilde}), i.e., the global phase,\footnote{The global phase is corrected slightly from that given previously~\cite{dArcy2001a}.} is not generally of interest.  Therefore, we absorb it into the transformed Floquet operator by defining $\tilde{F}_{n} \equiv \exp ( i Ma^{2} [3n^{2}-3n+1]T^3/6\hbar) \tilde{F}(nT,[n-1]T)$, such that
\begin{equation}
\tilde{F}_{n}=\exp\left(-\frac{i}{\hbar}\left[\frac{\hat{p}^{2}}{2M}T-\frac{\hat{p}a}{2}(2n-1)T^{2}\right]\right)\exp\left(i\phi_{d} \cos(K\hat{z}) \right).
\label{Floquet_derivation}
\end{equation}

Finally, we substitute the discrete and continuous components of $\hat{z}$ and $\hat{p}$ from Eq.\ (\ref{spatial_periodicity}) into Eq.\ (\ref{Floquet_derivation}).  A further simplification is then possible due to quasimomentum conservation (see Sec.\ \ref{sec:cons_quasimom}) \cite{Bach2005, Fishman2002, Fishman2003}.  When the transformed Floquet operator acts on a general quasi-momentum eigenstate $|\psi (\beta) \rangle = \sum_{k} c_{k}|k + \beta \rangle$, where the dimensionless momentum eigenkets $|k+\beta\rangle$ satisfy $\hat{p}|k+\beta\rangle = \hbar K(k+\beta) |k+\beta\rangle$ and $\langle k'+\beta' | k+ \beta\rangle= \delta_{kk'}\delta(\beta-\beta')$  \cite{Saunders2007}, the operator $\hat{\beta}$ can be replaced by its eigenvalue $\beta$.  Hence, the transformed Floquet operator, restricted to acting on a particular quasimomentum subspace \cite{Bach2005}, is
\begin{equation}
\begin{split}
\tilde{F}_n(\beta) =& \exp\left( -\frac{i}{\hbar} \left[
\frac{\hbar^2K^2}{2M}(\hat{k}+\beta)^{2}T-\pi\hbar \Omega (\hat{k}+\beta)(2n-1)\right] \right)  \\
& \times \exp\left( i\phi_{d} \cos (K\hat{z}) \right).
\label{Floquet_again}
\end{split}
\end{equation}

In Eq.\ (\ref{Floquet_again}), we have defined $\Omega \equiv KaT^{2}/2\pi$.  This parameter is a dimensionless rescaling of the effective gravitation acceleration $a=g-a_{\phi}$, in the frame comoving with the laser walking wave (see Sec.\ \ref{sec:walking_wave}).  In the context of quantum accelerator modes, $\Omega$ has been referred to as the unperturbed winding number \cite{Buchleitner2006}.  This is in analogy with the sine-circle map \cite{Schuster1995}, which can be considered a particular dissipative limiting case of the dynamics of the quantum $\delta$-kicked accelerator. 

\subsubsection{Quantum resonances and antiresonances \label{sec:quantum_resonances}}
 
The time evolution described by the transformed Floquet operator~(\ref{Floquet_again}) varies significantly depending on the system parameters.  In this paper, we investigate the effect of the effective gravity parameter $\Omega$ on quantum resonant dynamics in the strong-driving regime ($K^2 T\hbar \phi_d/M \geq1.5$).  In the $\delta$-kicked particle system (i.e., the $\delta$-kicked accelerator with $\Omega=0$), quantum resonances and antiresonances occur for pulse periodicities $T$ equal to integer multiples of the half-Talbot time $T_{T}/2$ \cite{Deng1999}, i.e.,
\begin{equation}
T =  \ell \frac{T_{T}}{2}= \ell \frac{2\pi M}{\hbar K^2}, 
\label{define_ell}
\end{equation}
where $\ell$ is a positive integer (see \cite{Saunders2007} and references therein).  In this paper we  consider in detail pulse periodicities given by Eq.\ (\ref{define_ell}) only.

The transformed Floquet operator (\ref{Floquet_again}), for the case where the pulse periodicity $T$ is given by Eq.\ (\ref{define_ell}), simplifies to 
\begin{equation}
\tilde{F}_{n} (\beta) = 
e^{ -i \pi \left[
\beta^{2}\ell
-\beta \Omega\left(2n-1\right)
\right] } e^{
-i \pi \left[ \hat{k}\ell +
2\hat{k}\beta\ell
-\hat{k} \Omega \left(2n-1\right)
\right]}
e^{i\phi_{d} \cos \hat{\theta}},
\label{Floquet_Talbot}
\end{equation}
where, because the eigenvalues of $\hat{k}$ are integer, we have used that $\exp(-i\ell \pi \hat{k}^2)=\exp(-i\ell \pi \hat{k})$ \cite{Bach2005,Fishman2002,Fishman2003}.   In the case where $T=\ell T_T/2$, the effective gravity parameter is $\Omega =  2\pi\ell^{2}a M ^{2}/\hbar^{2}K^{3}$.

The transformed Floquet operator (\ref{Floquet_Talbot}) can be written in a more concise way by defining 
\begin{equation} 
\label{Kfn}
 K \gamma_{n} \equiv \pi \left[ (1+2\beta)\ell - \Omega\left(2n-1\right) \right],
\end{equation}
i.e., 
\begin{equation}
\tilde{F}_{n} (\beta)= 
e^{ i [  (1+\beta)\pi  \ell - K \gamma_n ]\beta}
e^{-i K \gamma_n \hat{k}}
e^{ i\phi_{d} \cos \hat{\theta}}.
\label{eq:Floquet_last}
\end{equation}
The first term in the transformed Floquet operator~(\ref{eq:Floquet_last}) provides a quasimomentum dependent phase and the third term describes the momentum kick due to a laser field pulse.  Considering the second term in Eq.\ (\ref{eq:Floquet_last}) it is evident that, for an integer value of $\Omega$, $\exp(-iK\gamma_n\hat{k}) \rightarrow \exp(-iK\gamma_0 \hat{k})$ and the $n$-dependence of the transformed Floquet operator (\ref{eq:Floquet_last}) drops out.  Furthermore, when $\beta=0$, $K\gamma_{0}\rightarrow \pi(\ell + \Omega)$, and we find that integer changes in the value of $\Omega$ can be  equivalently described by modifying the number of half-Talbot times making up a kick period $T$, i.e., we can define an effective kick period $T=\ell'T_T/2$ from the integer $\ell'=\ell+\Omega$.  Recall that, for the $\delta$-kicked rotor where $\Omega=0$, even (odd) values of $\ell$ are associated with resonant (antiresonant) behaviour \cite{Saunders2007}.  Therefore, we expect that for a sufficiently narrow atomic momentum distribution centred around zero, quantum resonance (antiresonance) will be observed for even (odd) $\ell'$.  Indeed, a change in $\Omega$ by an odd integer value will change the system dynamics from resonant to antiresonant (or vice versa).

\subsubsection{Rational values of $\Omega$ \label{sec:rat_omega}}

The main focus of this paper is on rational values (particularly non-integer values) of $\Omega$, i.e., $\Omega = r/s$, where $r$ and $s$ are integers with no common factors and $s >0$.  To illustrate why these particular values are of interest it is useful to make one further transformation of the Hamiltonian (\ref{Hamiltonian_gauge}) into the frame where the atom cloud is on average stationary, i.e., we apply the unitary operator $\hat{U}_5 = \exp(-i[a\hat{p}t^{2}/2-Ma^{2}t^{3}/6]/\hbar)$ to the Hamiltonian (\ref{Hamiltonian_gauge}).  The transformed Hamiltonian is
\begin{equation}
\tilde{H}_{\rm stat} = \frac{ \hat{p}^2}{2M}-\hbar \phi_d \cos(K\hat{z}-n^2 \pi\Omega)\sum_{n=0}^{\infty} \delta (t-nT).
\label{Hstat}
\end{equation}

In this frame, we see that the effect of $\Omega$, when the $n$th kick is applied, is to shift the phase of the standing-wave potential by an amount $n^2 \pi \Omega$.  For the case where $\Omega = r/s$, this introduces an additional temporal periodicity into the system because the phase of the optical standing-wave pulses repeats identically with a period denoted by $n_T$ kicks.  The additional periodicity can equivalently be deduced from the Floquet operator (\ref{eq:Floquet_last}).  In Appendix \ref{Tperiodicity} we show that $n_T=s(1 + sr\bmod{2})$, i.e., $n_T=s$ for even $sr$, and $n_T=2s$ for odd $sr$.  The Hamiltonian (\ref{Hstat}) can then be rewritten as
\begin{equation}
\tilde{H}_{\rm stat} = \frac{ \hat{p}^2}{2M}-\hbar \phi_d 
\sum_{n'=0}^{n_{T}-1}\cos(K\hat{z}-n'^2 \pi r/s)\sum_{n=0}^{\infty} \delta (t-[n_{T}n+n']T),
\label{Hstat_rational}
\end{equation}
which applies for all possible values of $T$.

For the case where $\Omega$ is an even integer, $n_T=1$ and the standing-wave pulses are always in phase.  It follows that the atomic dynamics are unchanged from the $\Omega=0$ case.  When $\Omega$ is an odd integer, $n_T=2$ and from Eq.\ (\ref{Hstat_rational}) we find that consecutive optical pulses are exactly $\pi$ radians out of phase.  We primarily consider $T=\ell T_{T}/2$, in which case the dynamics for even $\Omega$ will always be either resonant or antiresonant depending on whether $\ell$ is even or odd, respectively.  Changes in $\Omega$ by an odd integer cause the dynamics to swap between resonant and antiresonant, as described in Sec.\ \ref{sec:quantum_resonances} and addressed further in Sec.\ \ref{sec:effect_omega}.  Fractional values of $\Omega$ lead to \textit{fractional resonances and antiresonances}.  Fractional resonances will be the main focus of this paper (see Sec.\ \ref{sec:fractional}).

\subsubsection{Momentum eigenstate evolution \label{sec:estate_evolve}}

It is convenient to first determine the time-evolution of the system when the centre-of-mass state of the atoms is initially prepared in the momentum eigenstate
\begin{equation}
| \Psi (t=0) \rangle = |k+\beta \rangle.
\label{init_state}
\end{equation}
Once the evolution of state (\ref{init_state}) is known, the evolution of any general state can be determined, as illustrated previously for the $\Omega=0$ case \cite{Saunders2007}.  

At time $t=nT$, the evolution of state (\ref{init_state}) can be determined by consecutively applying transformed Floquet operators defined by Eq.\ (\ref{eq:Floquet_last}), for each of the $n$ periods. More formally,  
\begin{equation}
| \Psi (t=nT) \rangle = {\tilde{\cal F}}_n(\beta) |k+\beta \rangle,
\label{final_state}
\end{equation}
where 
\begin{equation}
{\tilde{\cal F}}_n (\beta) = \tilde{F}_{n}(\beta)\tilde{F}_{n-1}(\beta)\tilde{F}_{n-2}(\beta) \dots \tilde{F}_{1}(\beta).
\label{time_ordering}
\end{equation}

Due to quasimomentum conservation (see Sec.\ \ref{sec:cons_quasimom}), Eq.\ (\ref{final_state}) can be expanded using momentum eigenstates with the initial quasimomentum $\beta$, i.e.,
\begin{equation}
| \Psi (t=nT) \rangle = \sum_{j=-\infty}^{\infty} c_{kj} (\beta,nT)|j+\beta \rangle,
\label{psi_nT}
\end{equation}
where the probability amplitudes $c_{kj} (\beta,nT)$ are given by
\begin{equation}
c_{kj} (\beta,nT) \delta(\beta-\beta') = \langle j+\beta' |\tilde{\cal{F}}_n(\beta)|k+\beta \rangle.
\label{mat_ele}
\end{equation}
Evaluation of the matrix elements (\ref{mat_ele}) closely follows our earlier derivation presented for the $\Omega=0$ case \cite{Saunders2007}, and is given in detail elsewhere \cite{Halkyard2008}.  In summary, 
\begin{equation}
c_{kj}(\beta,nT) = J_{j-k}(\omega) e^{i(j-k)\chi} e^{-2in\Upsilon k} e^{in^2\pi(k+\beta)\Omega} e^{-in\pi \beta^2 \ell},
\label{elements}
\end{equation}
where\footnote{The parameter $\nu$ in this paper is related to a previously defined parameter $\mu$ \cite{Saunders2007} according to $\nu=i\mu^*$.}
\begin{equation}
\omega e^{i\chi} = \phi_d \nu
\label{Eq:omega_nu}
\end{equation} 
for
\begin{equation}
\nu = ie^{-i(2n\Upsilon-n^2\pi\Omega)}\sum_{j=0}^{n-1} e^{i(2j\Upsilon-j^2\pi\Omega)},
\label{nu_eqn}
\end{equation}
and 
\begin{equation}
\Upsilon = \frac{1}{2}\pi (1+2\beta)\ell.
\label{upsilon}
\end{equation}
We will often specify the parameter dependence of $\nu$ by writing $\nu=\nu_{n,\ell}(\Omega,\beta)$. 

The sum in Eq.\ (\ref{nu_eqn}) can be evaluated analytically for particular choices of the dimensionless effective gravitational acceleration $\Omega$.  For integer values of $\Omega$, i.e., $\Omega = r_1$ \cite{Halkyard2008},
\begin{equation}
\nu_{n,\ell}(r_1,\beta) =  i e^{-i(n+1)(\Upsilon-\pi r_1/2)}\frac{\sin(n[\Upsilon - \pi r_1/2])}{\sin(\Upsilon - \pi r_1/2)}.
\label{int_nu}
\end{equation}
In Appendix \ref{appendix:find_nu} we consider half-integer values of $\Omega$, i.e., $\Omega = r_2/2$. We find that 
\begin{equation}
\nu_{n,\ell}(r_2/2,\beta) =2ie^{-i (\Upsilon [n+1]+\pi r_2/4)}\cos(\Upsilon - \pi r_2/4) \frac{\sin(n\Upsilon)}{\sin(2\Upsilon)}.
\label{half_int_nu}
\end{equation}
Note that Eq.\ (\ref{half_int_nu}) applies for even values of $n$ only.

In the $\beta=0$ subspace, where $\Upsilon\rightarrow \pi\ell/2$, $\nu$ can be evaluated analytically for $\Omega=1/s$, at kick values $n$ which are specific multiples of $s$ \cite{Halkyard2008}.  For $s(1-\ell)$ even, and $n$ a multiple of $s$,
\begin{equation}
\nu_{n,\ell}(1/s,0) = ie^{-i\pi[n\ell-n^2\Omega-(1-s\ell^{2})/4]} 
\frac{n}{\sqrt{s}}.
\label{nu_1os}
\end{equation} 
For $s(1-\ell)$ even or odd, and $n$ a multiple of $2s$,
\begin{equation}
\nu_{n,\ell}(1/s,0)= 
ie^{-i\pi[n\ell-n^2\Omega-(1-s\ell^{2})/4]} \frac{n[1+(-1)^{s(1-\ell)}]}{2\sqrt{s}}.
\label{nu_1osanti}
\end{equation} 
From Eq.\ ({\ref{nu_1osanti}}) two main classes of behaviour can be identified.  For odd values of $s(1-\ell)$, Eq.\ (\ref{nu_1osanti}) collapses to zero, i.e., the initial state is reconstructed every $2s$ kicks [see Eqs.\ (\ref{psi_nT}) and (\ref{elements}) with $\omega=0$].  We refer to this temporal reconstruction with period $2s$ as \textit{fractional antiresonant} behaviour.  In contrast, for even values of $s(1-\ell)$, Eq.\ (\ref{nu_1osanti}) reduces to Eq.\ (\ref{nu_1os}), and the magnitude of $\nu$ increases linearly with $n$ at a rate proportional to $1/\sqrt{s}$.  This leads to \textit{fractional resonant} behaviour, where energy is transferred from the laser field to the atom cloud in quasi-periodic bursts.  Fractional resonances are the main focus of this paper and are discussed in detail in Secs.\ \ref{sec:fractional} and \ref{sec:temperature}.

As in our previous work \cite{Saunders2007}, we study the evolution of momentum moments as a useful way to characterize the atomic dynamics.  If the state of the atoms is initially prepared in the momentum eigenstate $|k+\beta\rangle$, the $q$th-order momentum moment after time $nT$ is given by
\begin{equation}
\begin{split}
\langle \hat{p}^q \rangle_n &=   (\hbar K)^q \sum_{j=-\infty}^{\infty} |c_{kj}(\beta,nT)|^2(j+\beta)^q, 
\end{split}
\label{mom_evolve_zero}
\end{equation}
where $\omega$ is determined by substituting $\nu_{n,\ell}(\Omega,\beta)$, for example as given by Eqs.\ (\ref{int_nu}), (\ref{half_int_nu}), (\ref{nu_1os}), or (\ref{nu_1osanti}), into Eq.\ (\ref{Eq:omega_nu}).

\subsubsection{Evolution of an incoherent mixture}

To generalize the above treatment to the case of a cold thermal atomic cloud, we consider an incoherent mixture of plane waves with a Gaussian initial momentum distribution $D(p)=D_k(\beta)/\hbar K$, i.e., 
\begin{equation}
D_k(\beta) = \frac{1}{w \sqrt{2\pi}} \exp\left( \frac{-[k+\beta]^2}{2w^2}\right).
\label{eq:Gaussian}
\end{equation}
This corresponds to a Maxwell-Boltzmann distribution for free particles with temperature $\mathcal{T}_w=\hbar^{2}K^{2}w^{2}/Mk_{B}$.

The initial density operator for the system is taken to be  
\begin{equation}
\hat{\rho} =\hbar K \int dp|p\rangle
D(p)\langle p|
= \int_{-1/2}^{1/2} d\beta \sum_{k=-\infty}^{\infty}
|k+\beta\rangle
D_{k}(\beta)\langle k+\beta|.
\end{equation}
Evolving the density operator using Eq.\ (\ref{psi_nT}), and considering the diagonal elements, we find that after time $nT$ the momentum distribution evolves according to \cite{Saunders2007}
\begin{equation}
D_k(\beta,t=nT)  = \sum_{j=-\infty}^{\infty}  |c_{jk}(\beta,nT)|^2
D_{j}(\beta),
\label{eq:mom_dist}
\end{equation}
and the $q$th-order momentum moment is given by
\begin{equation}
\left\langle  \hat{p} ^q \right\rangle_n=
(\hbar K)^q \int_{-1/2}^{1/2} d\beta
\sum_{j,k=-\infty}^{\infty}
\left|c_{kj}(\beta,nT) \right|^2
D_{k}(\beta)(j+\beta)^{q}.
\label{eq:mom_mom_dist}
\end{equation}

\subsubsection{Numerical implementation}

The numerical results presented in this paper are generated using a previously described Monte Carlo approach \cite{Saunders2007}.  The initial condition consists of $\cal{N}$ plane waves distributed in momentum according to $D_k(\beta)$ of Eq.\ (\ref{eq:Gaussian}).  The time evolution of each momentum eigenstate is evaluated by applying the Floquet operator (\ref{Floquet_again}), which is represented in a momentum basis \cite{dArcy2001a}.  The final momentum distribution [see Eq.\ (\ref{eq:mom_dist})] is constructed by averaging the distributions resulting from each of the individual plane-wave time evolutions and collecting the data into a finite number of equal width momentum-space bins.  The resolution of the averaged atomic momentum distribution can be improved by increasing the number of bins per unit momentum $\hbar K$.

\section{Fractional resonances\label{sec:fractional}}

\subsection{Fractional resonances in the $\beta=0$ subspace\label{sec:fractional_beta_zero}}

\subsubsection{Fractional resonances with rational $\Omega$}

We first consider the dynamics of the atom-optical $\delta$-kicked accelerator when the initial state of the system is a zero momentum eigenstate, i.e., the dynamics are confined to the $\beta=0$ subspace.  Figure \ref{fig:ultracold_ell2} shows the resulting momentum distributions, and corresponding second- and fourth-order momentum moments, when the pulse periodicity is equal to the Talbot time $T_T$ ($\ell=2$).  We have chosen $\phi_{d}=0.8\pi$ as an illustrative value typical of recent experiments \cite{Oberthaler1999,Godun2000,dArcy2001a,dArcy2001b,Schlunk2003a,Schlunk2003b,dArcy2003,Ma2004,Buchleitner2006}.

As a basis for comparison with the fractional quantum resonances, we show the known kicked-rotor behaviour of the $\Omega=0$ case [see Figs.\ \ref{fig:ultracold_ell2}(a) and \ref{fig:ultracold_ell2}(b)], i.e., a quantum resonance is observed and the ballistic expansion of the atom cloud is characterized by quadratic growth in the second order momentum moment.  This growth is described by $\langle\hat{p}^2 \rangle_n=\hbar^2 K^2 \phi_d^2 n^2/2$, and is indicated by a solid line in Fig.\ \ref{fig:ultracold_ell2}(b) (see \cite{Saunders2007} and references therein).  

The quantum resonant dynamics are highly sensitive to the value of $\Omega$, and, as predicted in Sec.\ \ref{sec:quantum_resonances}, quantum antiresonance is observed for $\Omega=1$ [see Figs.\ \ref{fig:ultracold_ell2}(i) and \ref{fig:ultracold_ell2}(j)].  Quantum antiresonance is characterized here by reconstruction of the initial state every second kick, and the second- and fourth-order momentum moments oscillate accordingly (i.e., between the $n=0$ and $n=1$ values of Eqs.\ (\ref{eq:p2_res}) and (\ref{eq:p4_res}), respectively \cite{Saunders2007}).

When the effective gravity is chosen such that $\Omega=1/s$, where $s(1-\ell)$ is even (see Sec.\ \ref{sec:estate_evolve}), we observe fractional resonances [see Figs.\ \ref{fig:ultracold_ell2}(c), \ref{fig:ultracold_ell2}(e), and \ref{fig:ultracold_ell2}(g)].  At a fractional resonance, the energy transfer to the system is less efficient compared with the pure resonant case ($\Omega=0$).  The atomic momentum distribution expands in bursts of period $n_T=s$, and the quasi-periodic nature of the energy transfer is directly observable in the evolution of the second-order momentum moment [see Figs.\ \ref{fig:ultracold_ell2}(d), \ref{fig:ultracold_ell2}(f), and \ref{fig:ultracold_ell2}(h)].  The  quasi-periodic behavior arises because, in the frame where the atom cloud is on average stationary, the phase of the optical standing-wave pulses varies from kick to kick with period $n_T=s (1 + sr\bmod{2})$, as discussed in Sec.\ \ref{sec:rat_omega}.  
 
For comparison, Figs.\ \ref{fig:ultracold_ell2}(k) and \ref{fig:ultracold_ell2}(l) correspond to an irrational value of $\Omega$.  In this case, quasi-periodic behaviour is not observed.

\subsubsection{Momentum moment evolution \label{sec:evolution_moments_ultracold}}

For a zero momentum eigenstate, the dynamical evolution of the system is well characterized by the evolution of the second-order momentum moment, which, for $\Omega=1/s$, even $s(1-\ell)$, and $n$ an integer multiple of $s$, is given by \cite{Halkyard2008}
\begin{equation}
\langle \hat{p}^2\rangle_{n} = (\hbar K)^2
 \frac{\phi_d^2 n^2}{2s}.
\label{eq:p2_res}
\end{equation}
Equation (\ref{eq:p2_res}) describes quadratic growth in the second-order momentum moment (and therefore the kinetic energy) with a growth rate inversely proportional to $s$.\footnote{For odd values of $s(1-\ell)$ fractional antiresonances occur and all the momentum moments return to their zero initial value every 2s kicks [see Eq.\ (\ref{nu_1osanti})].}

\begin{figure}[tbp]
\includegraphics[width=8.5cm]{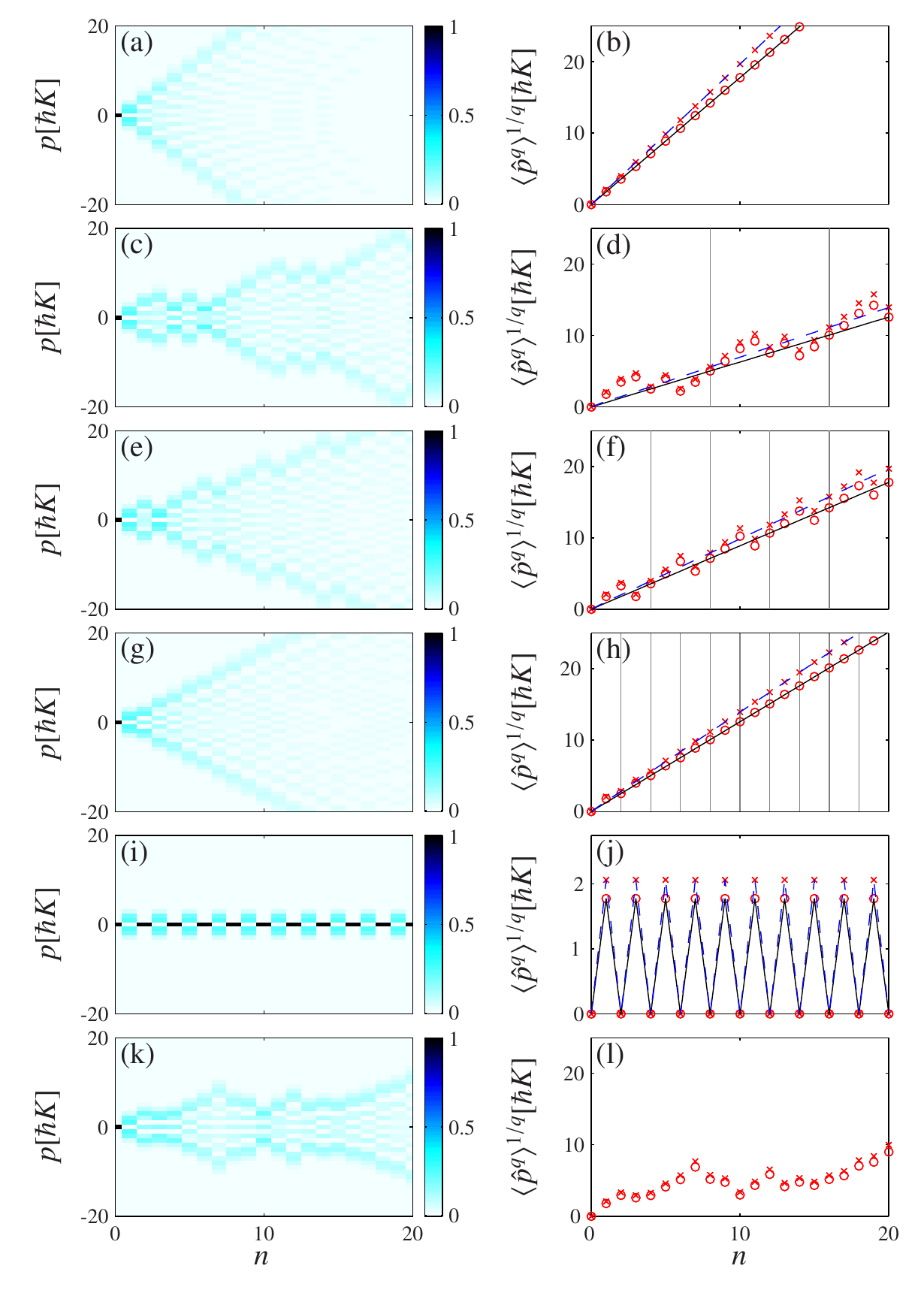}
\caption{(Colour online) (Left-hand panel) Momentum distributions with a resolution of $\hbar K$, and (right-hand panel) momentum moments of order ({\color{red}$\circ$}) $q=2$ and ({\color{red}$\times$}) $q=4$, for a $\delta$-kicked accelerator.  The initial condition is a zero-momentum eigenstate and parameters are $\mathcal{N}=1$, $T = T_T$ ($\ell=2$), $\phi_d = 0.8\pi$, and
(a), (b) $\Omega = 0$; 
(c), (d) $\Omega = 1/8$; 
(e), (f) $\Omega = 1/4$; 
(g), (h) $\Omega = 1/2$; 
(i), (j) $\Omega = 1$; and
(k), (l) $\Omega = (1+\sqrt{5})/2$.
In the right-hand panel the solid and dashed lines correspond to Eqs.\ (\ref{eq:p2_res}) and (\ref{eq:p4_res}), respectively.  The vertical lines in (d), (f), and (h), indicate where $n$ is an integer multiple of $s$ (as taken from $\Omega=1/s$).}
\label{fig:ultracold_ell2}
\end{figure}

In the same parameter regime, the fourth-order momentum moment evolution for an initial zero momentum eigenstate is \cite{Halkyard2008}
\begin{equation}
\langle \hat{p}^4 \rangle_{n}
= (\hbar K)^4 \left(\frac{3 \phi_d^4 n^4}{8 s^2} + \frac{\phi_d^2 n^2}{2s}\right).
\label{eq:p4_res}
\end{equation}
To leading order, Eq.\ (\ref{eq:p4_res}) describes quartic growth with a rate inversely proportional to $s^{2}$.  Therefore, the second- and fourth-roots of the second- and fourth-order momentum moments, respectively, each grow (to leading order) linearly in $n$ at a rate proportional to $1/\sqrt{s}$, as shown in Figs.\ \ref{fig:ultracold_ell2}(b), \ref{fig:ultracold_ell2}(d), \ref{fig:ultracold_ell2}(f), and \ref{fig:ultracold_ell2}(h).  We observe that, when $n$ is not an integer multiple of $s$, the momentum moments oscillate periodically around the analytic predictions (\ref{eq:p2_res}) and  (\ref{eq:p4_res}).

For the infinitely narrow initial momentum distributions considered here, the evolution of the fourth-order momentum moment does not provide significant additional information over and above that extracted from the second-order momentum moment evolution (see Fig.\ \ref{fig:ultracold_ell2}).  However, as we show in Sec.\ \ref{sec:fractional_finite_t}, the situation is quite different for initially broad, thermal momentum distributions, so we have included a description of the fourth-order momentum moment dynamics here for completeness.

\begin{figure*}[tbp]
\includegraphics[width=17cm]{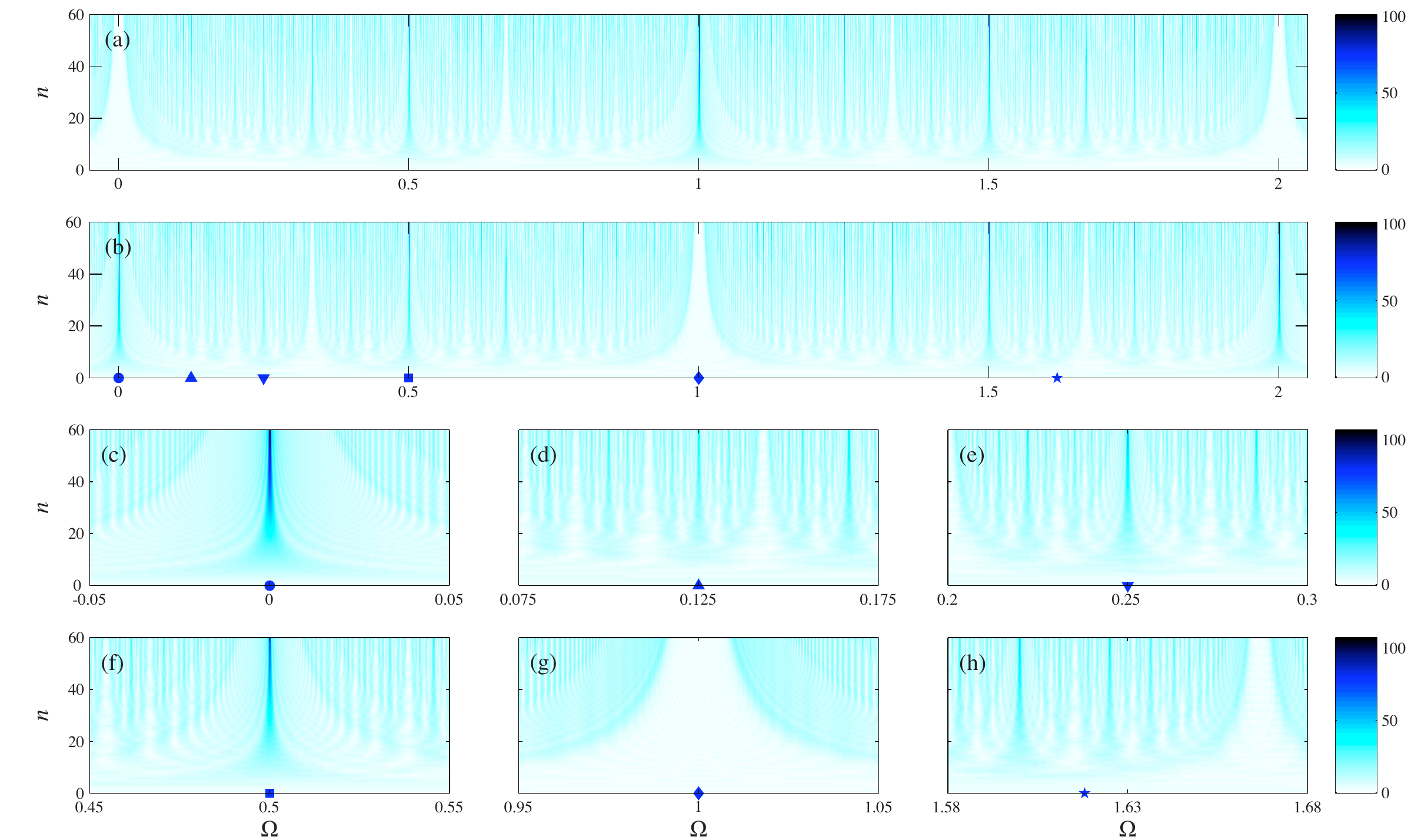}
\caption{(Colour online)
$\langle \hat{p}^{2} \rangle_{n}^{1/2}$ in units of $\hbar K$ for a $\delta$-kicked accelerator initially in a zero-momentum eigenstate.  Parameters are $\mathcal{N}=1$, $\phi_d = 0.8\pi$, and (a) $T=T_T/2$ ($\ell=1$) and (b) $T=T_T$ ($\ell=2$).
The markers highlight particular resonance features for $T=T_T$ ($\ell=2$), which are shown in more detail in
(c) $\Omega = 0$,
(d) $\Omega = 1/8$,
(e) $\Omega = 1/4$,
(f) $\Omega = 1/2$,
(g) $\Omega = 1$, and
(h) $\Omega = (1+\sqrt{5})/2$.}
\label{fig:omega_scan}
\end{figure*}

\subsubsection{Dependence on $\Omega$ of resonant features\label{sec:effect_omega}}

In Fig.\ \ref{fig:omega_scan} we study the evolution of the second-order momentum moment as $\Omega$ is varied continuously from $0$ to $2$ for both $T=T_{T}/2$ ($\ell=1$) and $T=T_{T}$ ($\ell=2$).  The initial condition is in every case a zero-momentum eigenstate.    We observe a rich structure of resonant and antiresonant features.  Resonant (antiresonant) features appear dark (pale) and become narrower and more prominent with increasing $n$.  The values of $\Omega$ used in Fig.\ \ref{fig:ultracold_ell2} are marked on Fig.\ \ref{fig:omega_scan}(b), and the regions immediately surrounding these values are shown in more detail in Figs.\ \ref{fig:omega_scan}(c) -- (h). 

In particular, we see from Figs.\ \ref{fig:omega_scan}(a) and \ref{fig:omega_scan}(b) that changing $\ell$ from $1$ to $2$ transposes the $\Omega$ dependence of the various resonant and antiresonant features by one, as can also be deduced from the structure of the transformed Floquet operator (\ref{eq:Floquet_last}) [see Sec.\ \ref{sec:quantum_resonances}].

\subsection{Fractional resonances for a finite temperature cloud \label{sec:fractional_finite_t}}

\subsubsection{Manifestation of fractional resonances at finite temperature}

In the case of a finite-temperature cloud, where the initial momentum distribution is described by the Gaussian\ (\ref{eq:Gaussian}), many quasimomentum subspaces are initially populated.  As previously discussed for $\Omega=0$ \cite{Saunders2007}, this results in some significant differences in the overall finite temperature dynamics compared with the dynamics of a system restricted to the $\beta=0$ subspace. 

\begin{figure}[tbp]
\includegraphics[width=8.5cm]{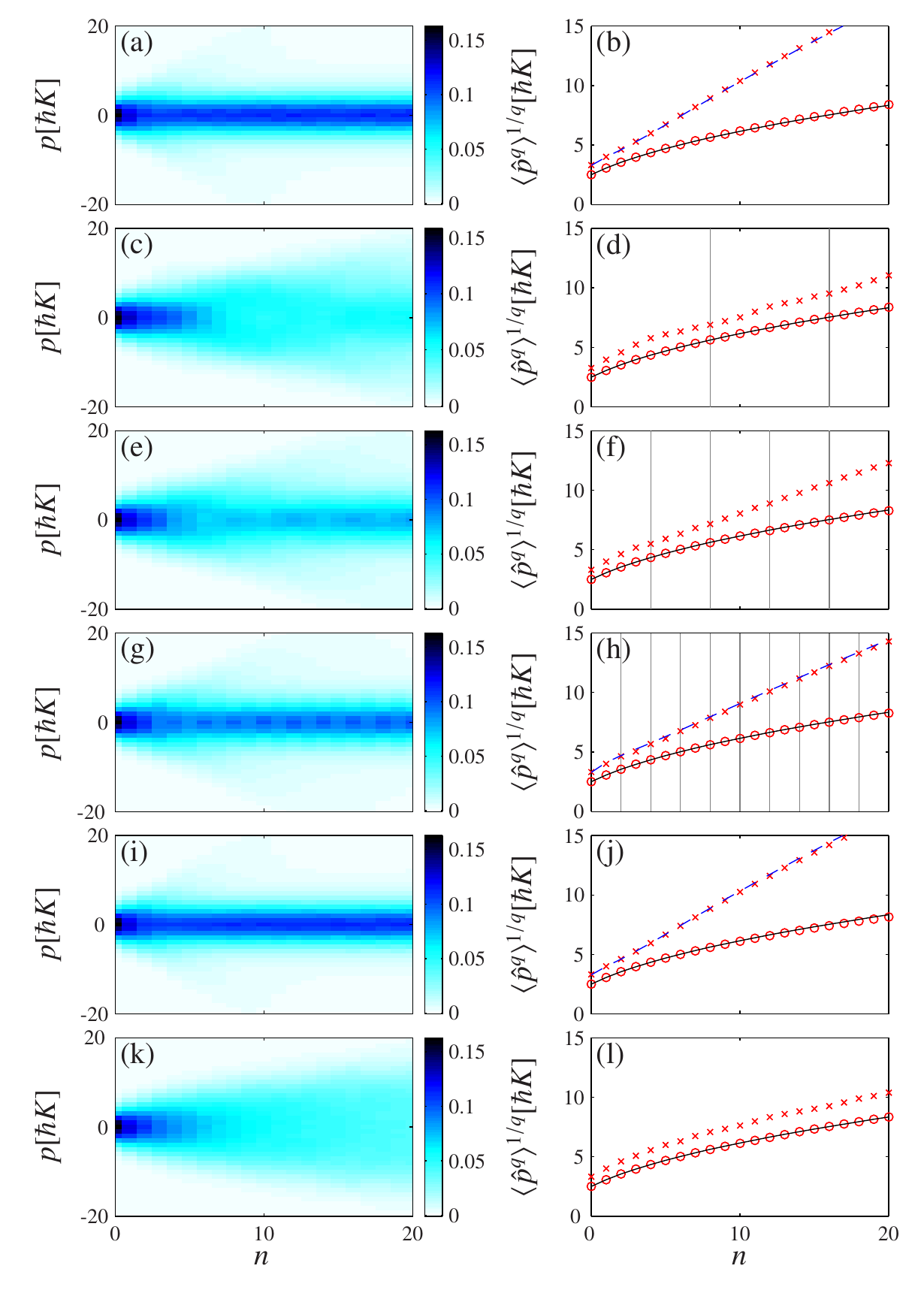}
\caption{
(Colour online) (Left-hand panel) Momentum distributions with a resolution of $\hbar K$, and (right-hand panel) momentum moments of order ({\color{red}$\circ$}) $q=2$ and ({\color{red}$\times$}) $q=4$, 
 for a $\delta$-kicked accelerator.  The initial atomic momenta are distributed according to Eq.\ (\ref{eq:Gaussian}) and parameters are  $\mathcal{N}=10000$, $w=2.5$, $T = T_T$ ($\ell=2$), $\phi_d = 0.8\pi$, and
(a), (b) $\Omega = 0$; 
(c),(d) $\Omega = 1/8$; 
(e), (f) $\Omega = 1/4$; 
(g), (h) $\Omega = 1/2$; 
(i), (j) $\Omega = 1$; and
(k), (l) $\Omega = (1+\sqrt{5})/2$.
In the right-hand panel, the solid lines correspond to Eq.\ (\ref{eq:p2_integer_hot}).
The dashed lines in (b) and (j) correspond to Eq.\ (\ref{eq:p4_integer_hot}), and the dashed line in (h) corresponds to Eq.\ (\ref{eq:p4_half_hot}). The vertical lines in (d), (f), and (h), indicate where $n$ is an integer multiple of $s$ (as taken from $\Omega=1/s$).
}
\label{fig:hot_ell2}
\end{figure}

Figure \ref{fig:hot_ell2} shows, as a function of kick number, the momentum distributions and second- and fourth-order momentum moments for the parameters used in Fig.\ \ref{fig:ultracold_ell2}, with the exception that in Fig.\ \ref{fig:hot_ell2} the initial condition in each case is a Gaussian momentum distribution with $w=2.5$  (corresponding to a temperature of approximately $5$ $\mu$K in the case of caesium \cite{dArcy2003}).  There are distinct differences in this case compared to the zero momentum eigenstate evolutions shown in Fig.\ \ref{fig:ultracold_ell2}.  For integer values of $\Omega$ \cite{Saunders2007,Halkyard2008}, considered for the sake of comparison with the fractional resonant cases,  pure resonant and antiresonant behavior occurs for the zero momentum eigenstate  [see Figs.\ \ref{fig:ultracold_ell2}(a) and \ref{fig:ultracold_ell2}(i)].  At finite temperature the momentum distribution behaves identically irrespective of whether $\Omega$ is even or odd [see Figs.\ \ref{fig:hot_ell2}(a) and \ref{fig:hot_ell2}(i)].  We observe that a small fraction of the cloud expands ballistically, but much of the distribution remains clustered near zero momentum.  This behavior is due to resonances and antiresonances occuring concurrently in a finite temperature cloud, and has been well described previously for $\Omega=0$ \cite{Saunders2007}.  

For $\Omega=1/s$, we again observe a small fraction of the cloud expanding ballistically at finite temperature [see Figs.\ \ref{fig:hot_ell2}(c),  \ref{fig:hot_ell2}(e), \ref{fig:hot_ell2}(g), and \ref{fig:hot_ell2}(i)].  However, as $s$ increases, the rate of ballistic expansion in the wings of the distribution is reduced.  This is characteristic of fractional resonances and was more clearly observable in the zero-mometum eigenstate evolutions shown in Fig.\ \ref{fig:ultracold_ell2}.  Also, with increasing $s$, the fraction of the cloud clustered near zero momentum becomes increasingly delocalized, as can be seen in Fig.\ \ref{fig:hot_ell2} after 20 kicks.  This indicates that, for rational values of $\Omega$, the antiresonances are not as effective at localizing the atom cloud.  A description of the fractional resonances at finite temperature is developed in Sec. \ref{sec:temperature}.

The evolutions of the second-order momentum moment are extremely similar for each value of $\Omega$ illustrated in Fig.\ \ref{fig:hot_ell2}.  We have found this to be a general property at finite temperature, which reflects the fact that resonant and antiresonant dynamics are occurring concurrently in the cloud.  This rules out the second-order momentum moment as a useful means of concisely distinguishing different fractional resonances from one another, or indeed from the non-resonant case shown in Figs.\ \ref{fig:hot_ell2}(k) and \ref{fig:hot_ell2}(l). 

In contrast, the fourth-order momentum moment evolutions remain comparatively distinct, as shown in the right-hand panel of Fig.\ \ref{fig:hot_ell2}.  We find that for larger values of $s$ (where $\Omega=1/s$), the fourth-order momentum moment increases at a slower rate.  This is a quantitative signature of the fact that as $s$ increases both the resonant and antiresonant dynamics become less efficient, and consequently the momentum distributions become less peaked.  The fourth-order momentum moment evolution is sensitive to this and provides a useful way to characterize the manifestation of fractional resonances at finite temperature.  However, we note that the fourth-order momentum moment evolutions for $\Omega=0$ [Fig.\ \ref{fig:hot_ell2}(b)] and $\Omega=1$ [Fig.\ \ref{fig:hot_ell2}(j)] are, like the second-order momentum moment evolutions, essentially indistinguishable.  

\subsubsection{Evolution of momentum moments \label{sec:moments_finite}}

Deriving analytic expressions for the momentum moment evolution of a finite temperature gas is more involved than arriving at the zero temperature results (\ref{eq:p2_res}) and (\ref{eq:p4_res}).  However, it is possible to derive such expressions for integer \cite{Halkyard2008} and half-integer values of $\Omega$.  In particular, we consider a thermal limit where $w$ is taken to be large for the initial Gaussian momentum distribution (\ref{eq:Gaussian}).

For integer values of $\Omega$, in the limit $w \gg 1/\sqrt{2}\pi \ell$ \cite{Halkyard2008},
\begin{equation}
\langle \hat{p}^2 \rangle_{n}= 
(\hbar K)^{2}
\left(w^2 + \frac{\phi_d^2 }{2} n\right),
\label{eq:p2_integer_hot}
\end{equation}
and 
\begin{equation}
\langle \hat{p}^4 \rangle_{n}
= 
(\hbar K)^4\left(\frac{\phi_d^4}{4}n^3 + \frac{\phi_d^4}{8}n + \frac{\phi_d^2}{2}n + 3w^2\phi_d^2 n + 3w^4\right).
\label{eq:p4_integer_hot}
\end{equation}
For half-integer values of $\Omega$ (see Appendix \ref{appendix:moments}), in the limit $w\gg1/2\sqrt{2} \pi \ell$, and with the restriction that $n$ is even, the evolution of the second-order momentum moment is again given by Eq.\ (\ref{eq:p2_integer_hot}), but the evolution of the fourth-order momentum moment is given by
\begin{equation}
\langle \hat{p}^4\rangle_{n}
= (\hbar K)^4
\left(\frac{\phi_d^4}{8}n^3 + \frac{5\phi_d^4}{8}n + \frac{\phi_d^2}{2}n + 3w^2\phi_d^2n + 3w^4\right).
\label{eq:p4_half_hot}
\end{equation}
Consequently, and as observed in Fig.\ \ref{fig:hot_ell2}, the system dynamics for integer and half-integer $\Omega$ values cannot be distinguished using $\langle \hat{p}^2 \rangle_{n}$.  However, the leading order term of the fourth-order momentum moment evolution is cubic in $n$, and the leading-order coefficient for half-integer $\Omega$ is exactly half the leading-order coefficient in the integer $\Omega$ case.  Therefore, it is possible to distinguish between fractional resonances with integer and half-integer values of $\Omega$ using $\langle \hat{p}^4\rangle_{n}$.

Our numerical calculations indicate that, for rational $\Omega=r/s$ and in the large-$w$ limit, $\langle \hat{p}^2 \rangle_{n}$ grows (on average) according to Eq.\ (\ref{eq:p2_integer_hot}), and $\langle \hat{p}^4 \rangle_{n}$ grows to leading order (on average) cubically with $n$ and with a growth rate proportional to $1/s$. We emphasize that while we have only shown these trends to be exactly true for $s=1$ and $s=2$, the generalization to higher values of $s$ is strongly supported by our numerical calculations.

\section{Effect of the initial atom cloud temperature on fractional resonances \label{sec:temperature}}

\subsection{Momentum dependence of fractional resonance features}

\subsubsection{Motivation \label{sec:motivation}}

In Sec.\ \ref{sec:fractional_beta_zero} we considered the dynamics of the $\delta$-kicked accelerator restricted to the $\beta=0$ subspace, and observed fractional resonances for $\Omega=1/s$ and $T=T_{T}$ ($\ell=2$).  We found that the fractional resonances could be characterized by quadratic growth of the second-order momentum moment at a rate inversely proportional to $s$, i.e., $\langle \hat{p}^{2}\rangle_n \propto n^2/s$ [see Eq.\ (\ref{eq:p2_res})].  Analogously, the fourth-order momentum moment evolves, to leading order, as $\langle \hat{p}^{4}\rangle_n \propto n^4/s^2$ [see Eq.\ (\ref{eq:p4_res})]. 

In contrast to the zero-temperature limit, at finite temperature (see Sec.\ \ref{sec:fractional_finite_t}) we found that the second-order momentum moment grows linearly with $n$, at a rate that appears to be essentially independent of $\Omega$, i.e, $\langle \hat{p}^{2}\rangle_n\propto n$, and that the fourth-order momentum moment appears to evolve, to leading order,  cubically as $\langle \hat{p}^{4}\rangle_n\propto n^3/s$.  

The differences observed between the zero-temperature limit and the thermal (large-$w$) limit arise because different quasimomentum subspaces evolve according to slightly different (i.e., $\beta$-dependent) transformed Floquet operators [see Eq.\ (\ref{eq:Floquet_last})].  For a broad initial momentum distribution, all quasimomentum subspaces are populated and the observed momentum moment dynamics result from the appropriate average over all the different $\beta$-subspace evolutions \cite{Saunders2007}.  It is therefore instructive to consider the dynamics of the different $\beta$ subspaces independently.

\subsubsection{Role of the quasimomentum $\beta$ \label{sec:quasimomentum}}

To better understand the $\beta$-dependence of the system evolution, we investigate the evolution of momentum eigenstates $|k+\beta\rangle$, where $k=0$ and $\beta \in [-1/2,1/2)$.  In particular, we consider the evolution of the momentum variance
\begin{equation}
\langle \langle \hat{p}^2 \rangle \rangle _n = \langle \hat{p}^2 \rangle_n - \langle \hat{p} \rangle_n^2.
\label{variance}
\end{equation}
For a momentum eigenstate, the variance displays clear signatures of the fractional resonances through the second-order momentum moment [see Sec.\ \ref{sec:fractional_beta_zero} and Fig.\ \ref{fig:ultracold_ell2}].  However, the variance is more useful in this context because it evolves independently of $k$ \cite{Saunders2007}, and its initial value is always zero irrespective $k$ and $\beta$. (The variance is introduced in a more general discussion of cumulants in Sec.\ \ref{sec:cumulant_background}.)

\begin{figure}[tbp]
\includegraphics[width=8.5cm]{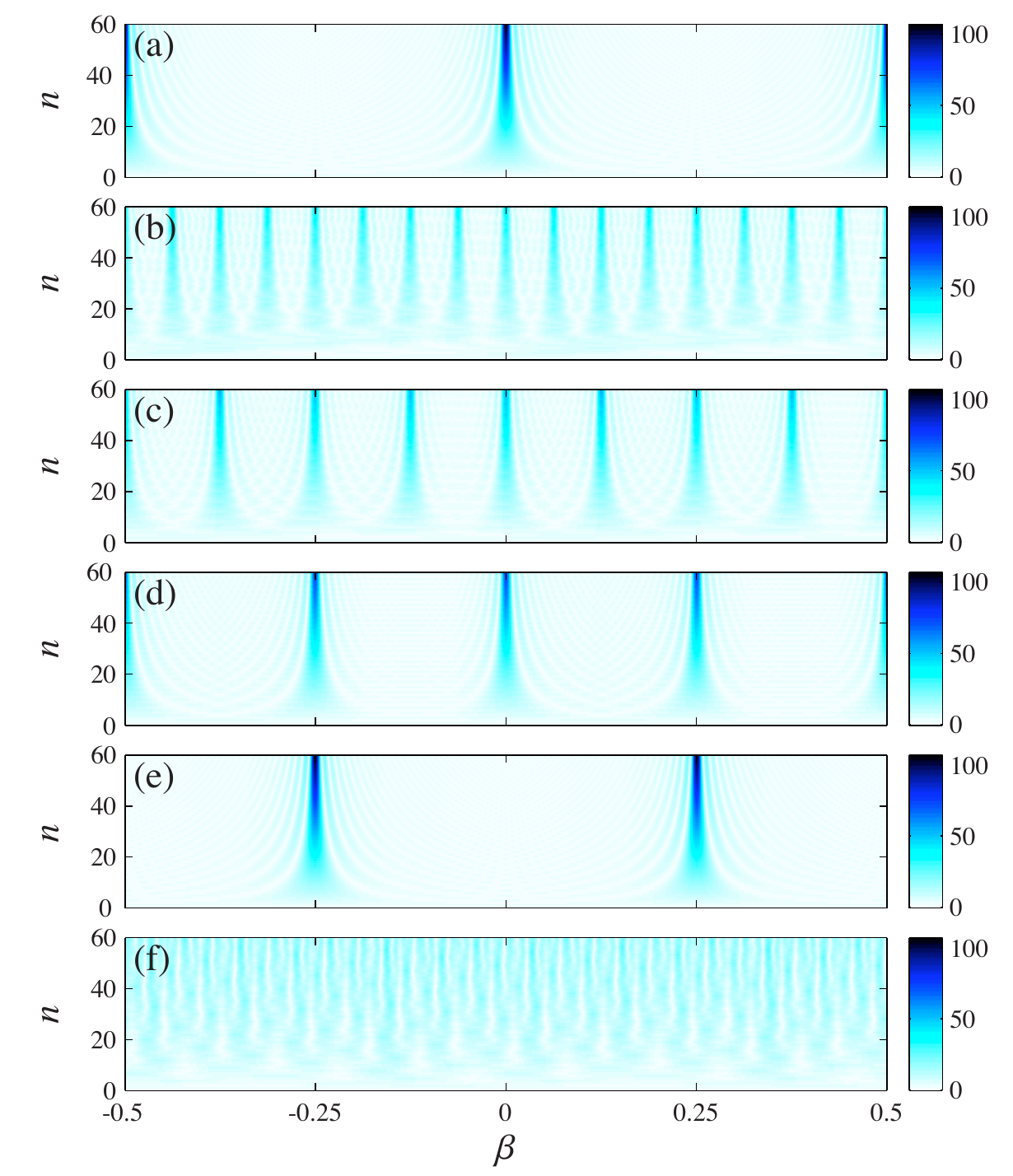}
\caption{
(Colour online) $\langle\langle\hat{p}^2\rangle\rangle_n^{1/2}$  in units of $\hbar K$ for a $\delta$-kicked accelerator where the system is initially prepared in the momentum eigenstate $|\beta\rangle$.  Parameters are $\mathcal{N}=1$, $\phi_d = 0.8\pi$, $T = T_T$ ($\ell=2$),  and 
(a) $\Omega = 0$,
(b) $\Omega = 1/8$,
(c) $\Omega = 1/4$,
(d) $\Omega = 1/2$,
(e) $\Omega = 1$, and
(f) $\Omega = (1+\sqrt{5})/2$.}
\label{fig:betascan}
\end{figure}

Figure \ref{fig:betascan} shows the quasimomentum dependence of the evolution of $\langle\langle \hat{p}^2 \rangle\rangle^{1/2}_{n}$, for $T=T_{T}$ ($\ell=2$), and for a variety of different values of $\Omega$.  In Fig.\ \ref{fig:betascan}(a), we show, for the sake of comparison, the known results for the $\delta$-kicked particle ($\Omega=0$): resonances occur for $\beta=0$ and $\beta=\pm 1/2$, and antiresonances occur for $\beta=\pm 1/4$ \cite{Saunders2007}.  For $\Omega=1$ [see Fig.\ \ref{fig:betascan}(e)], the resonance and antiresonance features are shifted in momentum by $1/4$ compared to the $\Omega=0$ case.

For $\Omega = 1/s$, Figs.\ \ref{fig:betascan}(b)--(e) show that increasing $s$ increases the density in $\beta$ of resonance and antiresonance features.  However, the resonances and antiresonances remain evenly spaced and, as we discuss in Sec.\ \ref{sec:res_width}, the width of the fractional resonances are independent of $s$.

In Fig.\ \ref{fig:betascan}(f) we show the quasimomentum dependence of the evolution of 
$\langle\langle \hat{p}^2 \rangle\rangle^{1/2}_{n}$ for an irrational value of $\Omega$.  We find that the resonance and antiresonance features observed for rational values become irregular and less well defined.

\subsubsection{Fractional resonance condition \label{sec:cond_frac_res}}

It is possible to quantify the $\beta$ separation between fractional resonances.  In the $\beta=0$ subspace, for $\Omega=1/s$, fractional resonances occur for even $s(1-\ell)$ (see Secs.\ \ref{sec:estate_evolve} and \ref{sec:fractional_beta_zero}).  In general, for $\Omega=r/s$, it can be shown that fractional resonances occur for even $s[r - (1+2\beta)\ell] $ \cite{Halkyard2008}.  Inverting this, we find that fractional resonances occur for  quasimomentum values 
\begin{equation}
\beta_m^{\rm FR} = \frac{r-\ell}{2\ell} - \frac{m}{\ell s},
\label{eq:resonance_condition}
\end{equation}
where $m$ is integer.  The fractional resonances are separated in momentum by $1/\ell s,$ breaking up momentum space into qualitatively similar (although not identical) regions.  The fractional resonances observed in Fig.\ \ref{fig:betascan} are consistent with Eq.\ (\ref{eq:resonance_condition}).  

\subsubsection{Fractional resonance width \label{sec:res_width}}

\begin{figure}[tbp]
\includegraphics[width=8.5cm]{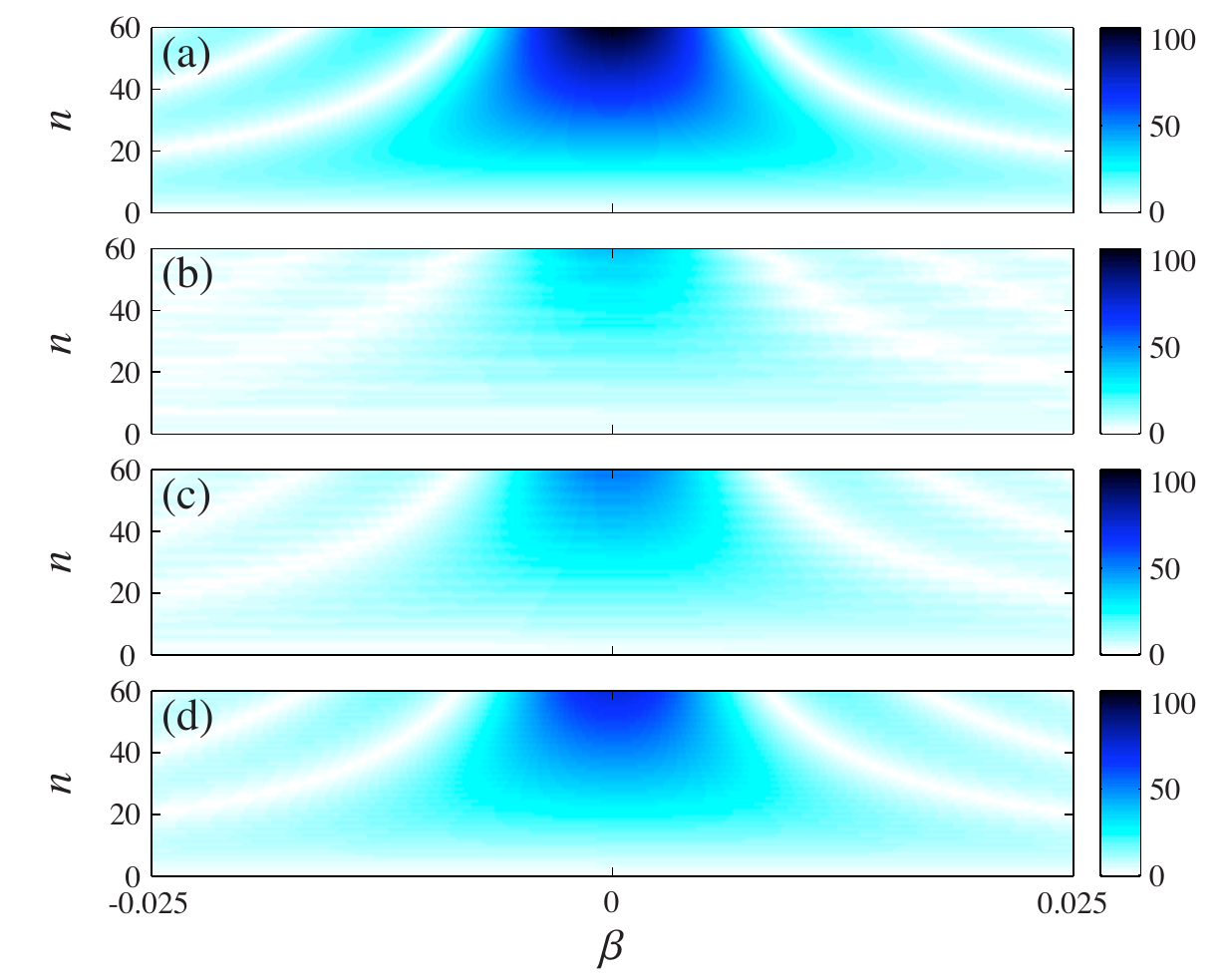}
\caption{
(Colour online) $\langle\langle\hat{p}^2\rangle\rangle_n^{1/2}$  in units of $\hbar K$ for a $\delta$-kicked accelerator where the system is initially prepared in the momentum eigenstate $|\beta\rangle$.  Parameters are $\mathcal{N}=1$, $\phi_d = 0.8\pi$, $T = T_T$ ($\ell=2$),  and 
(a) $\Omega = 0$,
(b) $\Omega = 1/8$,
(c) $\Omega = 1/4$, and
(d) $\Omega = 1/2$.}
\label{fig:betazoom}
\end{figure}

Fractional resonances have a momentum width that depends on $\ell$ and the kick number $n$.  As shown previously for the $\delta$-kicked particle \cite{Saunders2007}, an expression for the resonance width can be derived by considering antiresonance features in close proximity to the fractional resonances.  Figure\ \ref{fig:betazoom} shows a zoom in on the $\beta=0$ fractional resonances from Figs.\ \ref{fig:betascan}(a)--(d).  We observe hyperbolic curves of zero momentum variance which we call reconstruction loci \cite{Saunders2007}.  These are due to higher-order antiresonances which periodically reconstruct the initial state. 

The locations of the higher-order antiresonances that form the reconstruction loci can be determined for $s=1$ and $s=2$ using Eqs.\ (\ref{int_nu}) and (\ref{half_int_nu}), respectively.  The derivation is identical to that of the $\Omega=0$ case \cite{Saunders2007}, and we find that the reconstruction loci are described by $\beta =m/n\ell$, where $m$ is an integer.  The reconstruction loci closest to the resonance are then described by $\beta = \pm1/n\ell$.

We define the fractional resonance width as follows.  The hyperbolic curves lying halfway between the fractional resonance and the reconstruction loci are the sequence of points at which the state begins the process of reconstruction.  These transition loci are described by $\beta = \pm 1/2n\ell$.  The resonance width can then be defined as the $\beta$ separation between the two transition loci adjacent to the fractional resonance, i.e., \begin{equation}
\delta \beta_{\rm FR} = \frac{1}{n \ell}.
\label{res_width}
\end{equation}

The fractional resonance width\ (\ref{res_width}) is independent of $s$ for $s=1$ and $s=2$.  Our numerical calculations indicate that the width of the fractional resonances are, in general, independent of $s$, as illustrated in Fig.\ \ref{fig:betazoom}.

\subsection{Temperature dependence of the momentum moment evolution \label{sec:t_dep_moments}}

\begin{figure}[tbp]
\includegraphics[width=8.5cm]{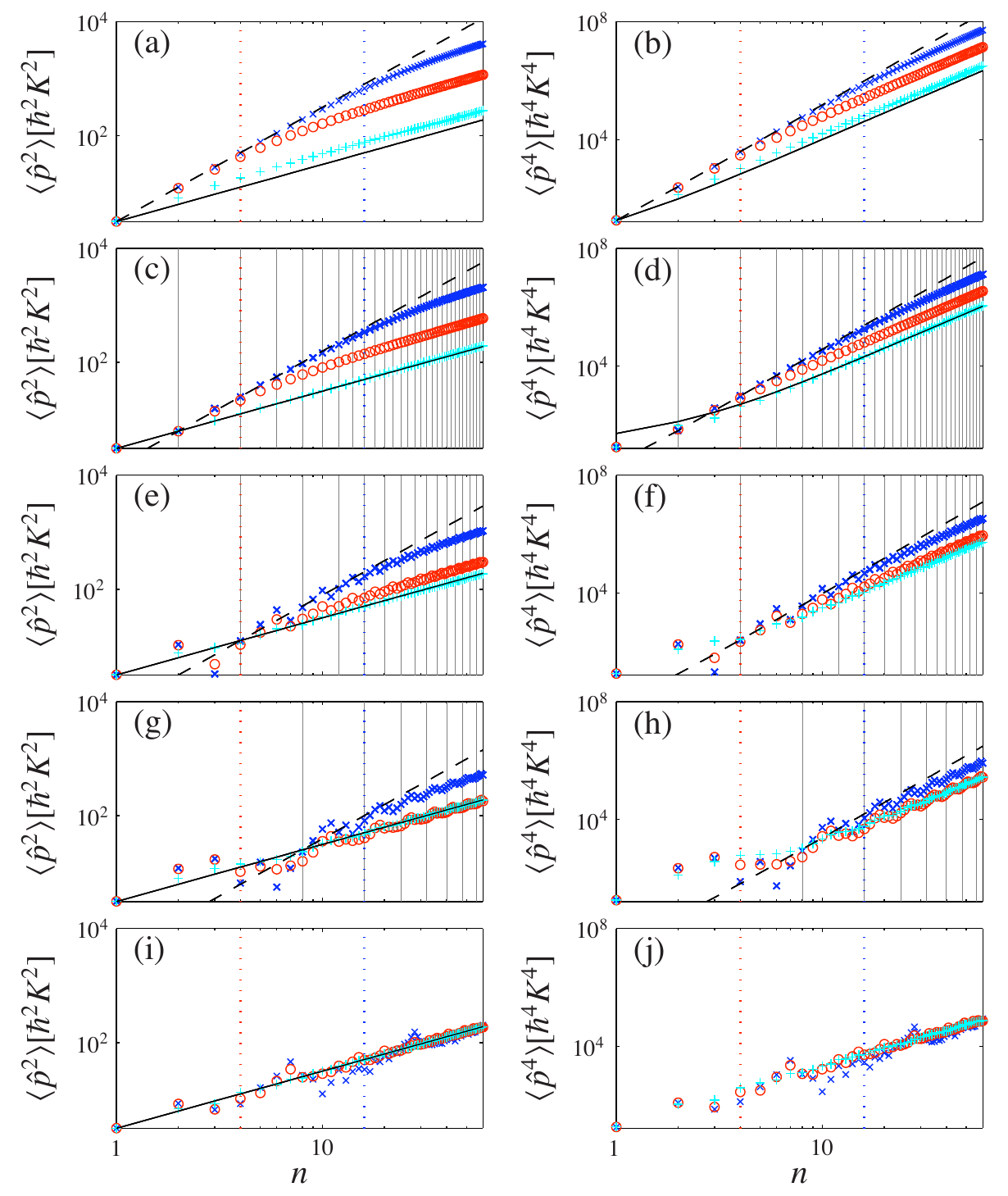}
\caption{
(Colour online)  (Left-hand panel) 
$\langle \hat{p}^{2}\rangle_n$ and (right-hand panel) $\langle \hat{p}^{4} \rangle_n$ for a Gaussian initial atomic momentum distribution with ({\color{blue} $\times$}) $w = 1/128$, ({\color{red} $\circ$}) $w = 1/32$, and ({\color{cyan} $+$}) $w = 1/8$.
Parameters are $\mathcal{N}=10000$, 
$T = T_T$ ($\ell=2$), $\phi_d = 0.8\pi$, and
(a), (b) $\Omega = 0$;
(c), (d) $\Omega = 1/2$;
(e), (f) $\Omega = 1/4$; 
(g), (h) $\Omega = 1/8$; and 
(i), (j) $\Omega = (1+\sqrt{5})/2$.
The dashed lines correspond to the $w=0$ analytic predictions [see (left-hand panel) Eq.\ (\ref{eq:p2_res}) and (right-hand panel) Eq.\ (\ref{eq:p4_res})].  The solid lines correspond to the large-$w$ limit lower bound [see (left-hand panel) Eq.\ (\ref{eq:p2_integer_hot}), (b) Eq.\ (\ref{eq:p4_integer_hot}), and (d) Eq.\ (\ref{eq:p4_half_hot}), all with $w=0$].  The vertical dotted lines indicate $n=n_{\rm FR}$ [see Eq.\ (\ref{n_FR})], and the solid vertical lines in (c)--(h) indicate where $n$ is an integer multiple of $s$ (as taken from $\Omega=1/s$).}
\label{fig:gravity_temp_moments}
\end{figure}

We investigate the effect of the initial atom cloud temperature by considering the evolution of the second- and fourth-order momentum moments for Gaussian initial atomic momentum distributions with different widths $w$ [see Eq.\ (\ref{eq:Gaussian})].  Figure \ref{fig:gravity_temp_moments} shows $\langle \hat{p}^{2}\rangle_n$ and $\langle \hat{p}^{4}\rangle_n$ for different values of $w$ and $\Omega$.  

As $w$ increases, for a given $\Omega$, the momentum moment evolutions make a gradual transition from the zero-temperature limit behavior discussed in Sec.\ \ref{sec:evolution_moments_ultracold}, towards the large-$w$ limit described in Sec.\ \ref{sec:moments_finite}.  The dashed lines in Fig.\ \ref{fig:gravity_temp_moments} indicate the zero-temperature limit, given by Eqs.\  (\ref{eq:p2_res}) and (\ref{eq:p4_res}) for the second- and fourth-order momentum moments, respectively.  The solid lines in the figure are \textit{lower bounds} for the momentum moment evolutions in the large-$w$ limit, i.e., they are given by the large-$w$ limit expressions in Sec.\ \ref{sec:moments_finite} evaluated with $w=0$.  For the second-order momentum moment, the solid lines correspond to Eq.\ (\ref{eq:p2_integer_hot}) with $w=0$.   Note that for finite $w$, the large-$w$ limit of $\langle \hat{p}^{2} \rangle_n$ will be larger than that for $w=0$, but will increase at the same rate.  For the fourth-order momentum moment evolution, the lower bounds depend on $\Omega$ and are defined by Eqs.\ (\ref{eq:p4_integer_hot}) and (\ref{eq:p4_half_hot}), each with $w=0$, for integer and half-integer $\Omega$, respectively.  We have not evaluated analytic expressions for the lower bounds for general rational $\Omega$, although Figs.\ \ref{fig:gravity_temp_moments}(f) and \ref{fig:gravity_temp_moments}(h) strongly indicate that similar large-$w$ limiting behavior exists.  In addition to the general growth trends of the momentum moments, the quasi-periodic behavior with period $n_T=s$ is still observable (see Sec.\ \ref{sec:rat_omega}).

The deviation of the momentum moment evolution from the zero-temperature limit occurs at a particular kick number $n_{\rm FR}$ (indicated by the vertical dotted lines in Fig.\ \ref{fig:gravity_temp_moments}).  As is well understood for the $\Omega=0$ case \cite{Saunders2007}, the energy transferred to the system becomes limited when the initial momentum width of the atom cloud becomes comparable with the momentum width of the resonance.  If the initial momentum width is large compared to the resonance width, antiresonances play a role in the system dynamics and the momentum moment evolution tends towards the thermal large-$w$ limit.  In the case of the fractional resonances the same arguments apply, and indeed the width of the fractional resonances are independent of $\Omega$, as described in Sec.\ \ref{sec:res_width}.  We define $n_{\rm FR}$ such that two standard deviations of the initial Gaussian momentum distribution lie within the resonance width $\delta \beta _{\rm FR}$ (see Sec.\ \ref{sec:res_width}), i.e.,
\begin{equation}
 n_{\rm FR} = \frac{1}{4w\ell}.
 \label{n_FR}
\end{equation}

In Fig.\ \ref{fig:gravity_temp_moments} we also find that as $s$ increases (where $\Omega=1/s$), the fractional resonances become increasingly sensitive to the initial temperature of the system, i.e., the momentum moments approach their large-$w$ limiting behavior for lower values of $w$.  This is most clearly observed in the second-order momentum moment, where the moment evolutions tend increasingly towards the solid line as $s$ increases. We can explain this in terms of the number of resonance features spanned by the initial atomic momentum distribution.  The fractional resonances are separated in momentum by $1/\ell s$ (see Sec.\ \ref{sec:cond_frac_res}).  Thus, we can define a thermalization width $w_{\rm th}$, above which the atomic momentum distribution initially spans more than a single resonance feature and the atomic dynamics reach a \textit{thermal} limit.  For $\Omega=1/s$, this would be manifest in, for example, linear rather than quadratic growth in $\langle \hat{p}^{2}\rangle_n$ with $n$, and cubic rather than quartic growth in $\langle \hat{p}^{4}\rangle_n$ (see Sec.\ \ref{sec:moments_finite}).  Quantitatively, we define $w_{\rm th}$ such that one standard deviation of the initial Gaussian momentum distribution lies within the fractional resonance separation, i.e.,
\begin{equation}
w_{\rm th} = \frac{1}{2\ell s}.
\label{w_th}
\end{equation}
Equation\ (\ref{w_th}) is consistent with the large-$w$ limit described in Sec.\ \ref{sec:moments_finite} (and taken more formally in Appendix\ \ref{appendix:moments}), where terms involving $\exp (-\pi^2 w^2/2w_{\rm th}^2)$ were taken to be negligible.
  
Finally, we note that if $\Omega$ is chosen to be irrational [see Figs.\ \ref{fig:gravity_temp_moments}(i) and \ref{fig:gravity_temp_moments}(j)], we observe that the second-order momentum moment grows linearly with $n$, and the fourth-order momentum moment grows quadratically.  The data becomes smoother as $w$ increases, but the growth rate appears to be largely independent of $w$.

\subsection{Momentum cumulant evolution}

\subsubsection{Motivation and background \label{sec:cumulant_background}}

In Sec.\ \ref{sec:t_dep_moments} we found that the evolution of the momentum moments depends explicitly on $w$ [see Eqs.\ (\ref{eq:p2_integer_hot}), (\ref{eq:p4_integer_hot}), and (\ref{eq:p4_half_hot})].  For this reason, we defined a lower bound for the large-$w$ limit of the momentum moment evolutions.  As can be shown explicitly for integer \cite{Halkyard2008} and half-integer values of $\Omega$, this difficulty does not arise if we consider the change in the momentum \textit{cumulants} from their initial value. In this way it is possible to have a well defined high-temperature limit.

A $q$th-order momentum moment $\langle\hat{p}^q\rangle_n$ is dependent upon all moments up to order $(q-1)$. In particular, $\langle\hat{p}^2\rangle_n$ and $\langle\hat{p}^4\rangle_n$ are not independent quantities.  Using an iterative process, mutually independent cumulants $\langle\langle\hat{p}^q\rangle\rangle_n$ can be constructed from the moments $\langle\hat{p}\rangle_{n} , \langle\hat{p}^2\rangle_n,   \dots,  \langle\hat{p}^q\rangle_{n}$  \cite{Bach2005,Gardiner2004,Fricke1996,Kohler2002}.  The first-order cumulant $\langle\langle\hat{p}\rangle\rangle_n$ is simply the mean $\langle\hat{p}\rangle_n$.  The second-order cumulant $\langle\langle\hat{p}^2\rangle\rangle_n$ is the variance [as defined in Eq.\ (\ref{variance})].  The fourth-order momentum cumulant, the \textit{kurtosis\/}, is 
\begin{equation}
\langle\langle\hat{p}^4\rangle\rangle = \langle\hat{p}^4\rangle
- 4 \langle\hat{p}^3\rangle \langle\hat{p}\rangle
+12 \langle\hat{p}^2\rangle \langle\hat{p}\rangle^{2}
- 3\langle\hat{p}^2\rangle^2
- 6 \langle\hat{p}\rangle^{4},
\label{eq:cumulant_long}
\end{equation}
where we have dropped the $n$ subscripts for brevity. 

For a symmetric momentum distribution, the odd-ordered momentum moments are identically zero, and the second- and fourth-order cumulants simplify to
\begin{align}
\langle\langle\hat{p}^2\rangle\rangle_n =& \langle\hat{p}^2\rangle_n, \label{eq:c2_definition}\\
\langle\langle\hat{p}^4\rangle\rangle_n =& \langle\hat{p}^4\rangle_n - 3\langle\hat{p}^2\rangle_n^2.
\label{eq:c4_definition}
\end{align}
If the momentum distribution is symmetric initially, one can show that, for $\Omega=r/s$, it must evolve to a symmetric distribution at intervals of $s$ kicks \cite{Halkyard2008}.  One can therefore infer (and indeed observes) that the odd moments should be bounded in value, and remain relatively insignificant in comparison to the rapidly growing even moments.  We therefore, at times, make use of Eq.\ (\ref{eq:c2_definition}) and Eq.\ (\ref{eq:c4_definition}) even when their use is not fully justified.  Numerically, the fact that we employ a Monte-Carlo method yields small, although in general non-zero, odd moments.  Therefore, in numerical calculations we evaluate the momentum moments using Eq.\ (\ref{eq:mom_mom_dist}), and then determine the cumulants from the moments using Eqs.\ (\ref{variance}) and (\ref{eq:cumulant_long}).  

\subsubsection{Zero-temperature limit}

\begin{figure}[tbp]
\includegraphics[width=8.5cm]{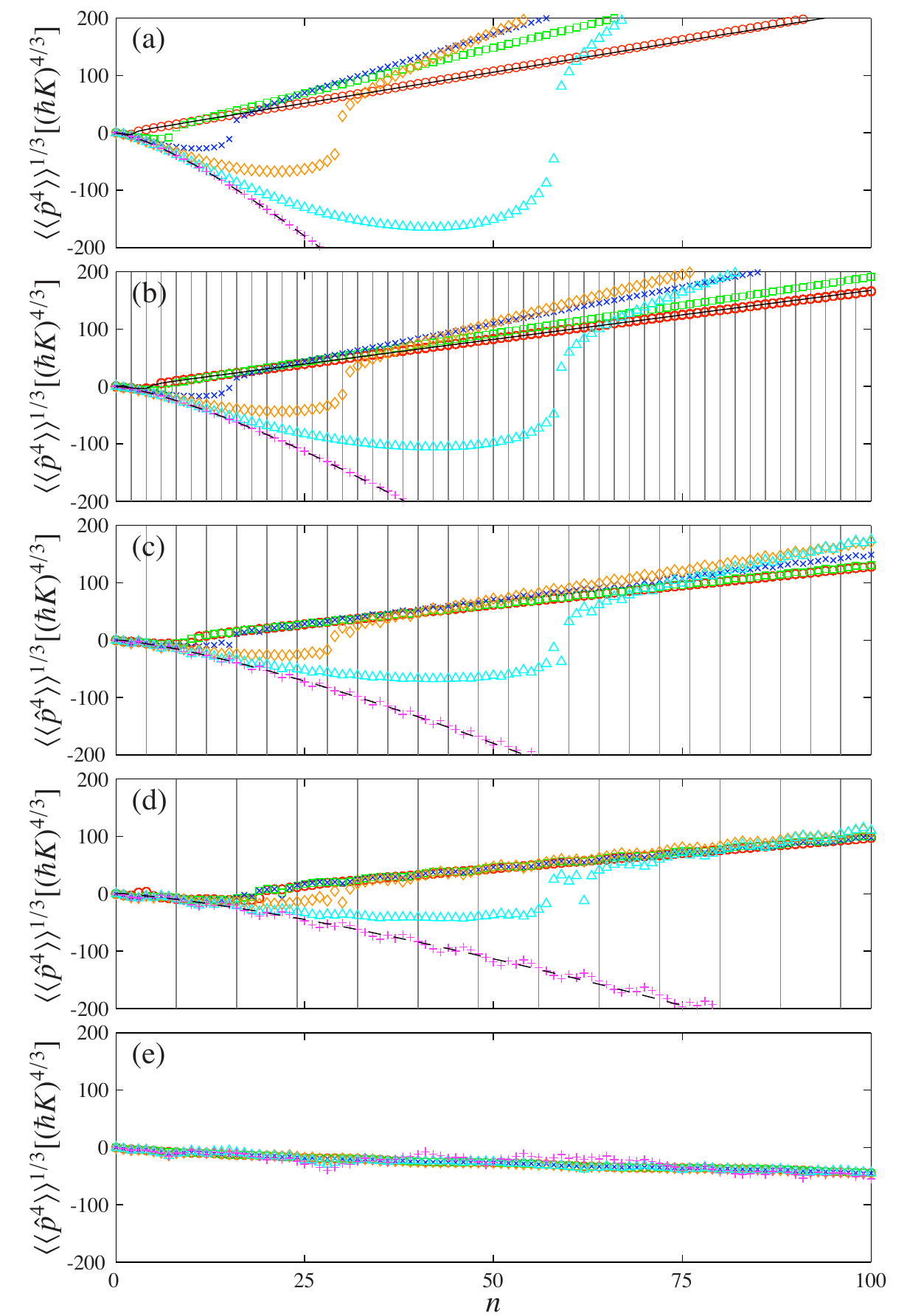}
\caption{
(Color online) $\langle \langle \hat{p}^4 \rangle\rangle_n^{1/3}$ for a $\delta$-kicked accelerator with Gaussian initial momentum distribution of standard deviation
({\color{red} $\circ$}) $w = 1/4$,
({\color{green} $\square$}) $w = 1/16$,
({\color{blue} $\times$}) $w = 1/32$,
({\color{orange} $\diamond$}) $w = 1/64$,
({\color{cyan} $\triangle$}) $w = 1/128$, and
({\color{magenta} $+$}) $w = 1/1024$.
Parameters are ${\cal N}=10000$, $T = T_T$ $(\ell=2)$, $\phi_d = 0.8\pi$, and 
(a) $\Omega = 0$,
(b) $\Omega = 1/2$,
(c) $\Omega = 1/4$,
(d) $\Omega = 1/8$, and
(e) $\Omega = (1+\sqrt{5})/2$.
The dashed lines correspond to Eq.\ (\ref{eq:c4_res}), and the solid lines correspond to (a) Eq.\ (\ref{eq:c4_i_hot}) and (b) Eq.\ (\ref{eq:c4_h_hot}).  The vertical lines in (b)--(d)  indicate where $n$ is an integer multiple of $s$ (as taken from $\Omega=1/s$).
}
\label{fig:cumulants}
\end{figure}

Before considering the high-temperature limit of the cumulant dynamics, it is first instructive to consider the zero-temperature limit where the initial state is a zero-momentum eigenstate.  The initial state is symmetric in $p$ so $\langle\langle\hat{p}^2\rangle\rangle_{n}=\langle\hat{p}^2\rangle_{n}$ is given by Eq.\ (\ref{eq:p2_res}) which is valid for $n$ an integer multiple of $s$ (where $\Omega=1/s$).  Substituting Eqs.\ (\ref{eq:p2_res}) and (\ref{eq:p4_res}) into Eq.\ (\ref{eq:c4_definition}) yields the known result \cite{Halkyard2008}
\begin{equation}
\langle\langle \hat{p}^4 \rangle\rangle_{n}
= (\hbar K)^4
\left(  - \frac{3 \phi_d^4 n^4}{8 s^2} +  \frac{\phi_d^2 n^2}{2s}\right),
\label{eq:c4_res}
\end{equation}
where again $n$ is an integer multiple of $s$.

\subsubsection{High-temperature limit}

For a Gaussian (and therefore symmetric) initial momentum distribution of standard deviation $w$,  by definition $\langle\langle \hat{p}^2 \rangle\rangle_{0}= (\hbar K)^2 w^{2}$ [see Eqs.\ (\ref{eq:c2_definition}) and (\ref{eq:p2_TERM1})]. In the thermal limit, where $w$ is taken to be large, the momentum variance is given by Eq.\ (\ref{eq:p2_integer_hot}).  Subtracting the initial value gives
\begin{equation}
\langle\langle \hat{p}^2 \rangle\rangle_{n}
- \langle\langle \hat{p}^2 \rangle\rangle_{0}
= (\hbar K)^2\frac{\phi_d^2 n}{2},
\label{eq:c2_hot}
\end{equation}
which is independent of $w$.  Note that Eq.\ (\ref{eq:c2_hot}) applies to both integer and half-integer values of $\Omega$ (although for half-integer values of $\Omega$ it applies only for even values of $n$).  Our numerical calculations strongly indicate that Eq.\ (\ref{eq:c2_hot}) should apply for all values of $\Omega$ (see Fig.\ \ref{fig:hot_ell2} and Sec.\ \ref{sec:moments_finite}).

A Gaussian distribution has no non-zero cumulants of order greater than two so, when considering an initially Gaussian momentum distribution, we have $\langle\langle \hat{p}^4 \rangle\rangle_{0}=0$.  For integer values of $\Omega$, we substitute Eqs.\ (\ref{eq:p2_integer_hot}) and  (\ref{eq:p4_integer_hot}) into Eq.\ (\ref{eq:c4_definition}) and find that
\begin{equation}
\langle\langle \hat{p}^4 \rangle\rangle_{n} 
= (\hbar K)^4\left(\frac{\phi_d^4}{4}n^3 - \frac{3\phi_d^4 n^2}{4} + \frac{\phi_d^4}{8}n + \frac{\phi_d^2}{2}n\right).
\label{eq:c4_i_hot}
\end{equation}
Equation (\ref{eq:c4_i_hot}) is also independent of $w$. An analogous expression for half-integer values of $\Omega$ can be found by substituting Eqs.\ (\ref{eq:p2_integer_hot}) and (\ref{eq:p4_half_hot}) into Eq.\ (\ref{eq:c4_definition}) to yield
\begin{equation}
\langle\langle \hat{p}^{4}\rangle\rangle_{n}
= (\hbar K)^4 \left(\frac{\phi_d^4}{8}n^3 - \frac{3\phi_d^4 n^2}{4} + \frac{5\phi_d^4}{8}n + \frac{\phi_d^2}{2}n\right),
\label{eq:c4_h_hot}
\end{equation}
which is valid for even $n$.  Equations (\ref{eq:c4_i_hot}) and (\ref{eq:c4_h_hot}) show that the cubic leading-order cumulant behavior, and its coefficient, is unchanged compared to the corresponding momentum moments [see Eqs.\ (\ref{eq:p4_integer_hot}) and (\ref{eq:p4_half_hot})].  For general rational values of $\Omega$, we expect that the change in the momentum cumulants, compared to their initial values, will again be independent of $w$, for $w$ sufficiently large. 

\subsubsection{Temperature dependence of the momentum cumulant evolution}

The second-order momentum cumulant evolution is essentially described by the second-order momentum moment evolution [see Eq.\ (\ref{variance}) with negligible $\langle \hat{p}\rangle_n$].  Therefore, the temperature dependence of the second-order momentum cumulant evolution does not provide any further information than that presented for the second-order momentum moment evolution in Sec.\ \ref{sec:t_dep_moments}.  In this section we concentrate on the temperature dependence of the fourth-order momentum cumulant evolution.

Figure\ \ref{fig:cumulants} shows the evolution of the cube-root of the fourth-order momentum cumulant for different values of $w$ and $\Omega$.  The fourth-order momentum cumulant initially evolves according to the zero-temperature limit, indicated by the dashed lines in the figure, and is negative indicating a broad momentum distribution characteristic of quantum resonance phenomena.  At the kick number $n=n_{\rm FR}$ [see Eq.\ (\ref{n_FR})], the cumulants deviate from the zero-temperature limit and eventually become positive, indicating a sharply peaked momentum distribution which is caused by antiresonances playing a significant role in the dynamics.  For irrational values of $\Omega$ [Figs.\ \ref{fig:cumulants}(e)], the absence of resonance and antiresonance features means that the fourth-order cumulant remains negative.

For sufficiently large values of $w$, the fourth-order momentum cumulant tends to the thermal limit, indicated by the solid lines in Figs.\ \ref{fig:cumulants}(a) and \ref{fig:cumulants}(b).  In the thermal limit, the fourth-order momentum cumulant evolution is independent of $w$ and is characterized by cubic growth with $n$ [see Eqs.\ (\ref{eq:c4_i_hot}) and (\ref{eq:c4_h_hot})].  The growth rate is smaller for higher values of $s$ (where $\Omega=1/s$), indicating the less peaked momentum distributions of higher-ordered fractional resonances (see Fig.\ \ref{fig:hot_ell2}).  

\begin{figure}[tbp]
\includegraphics[width=8.5cm]{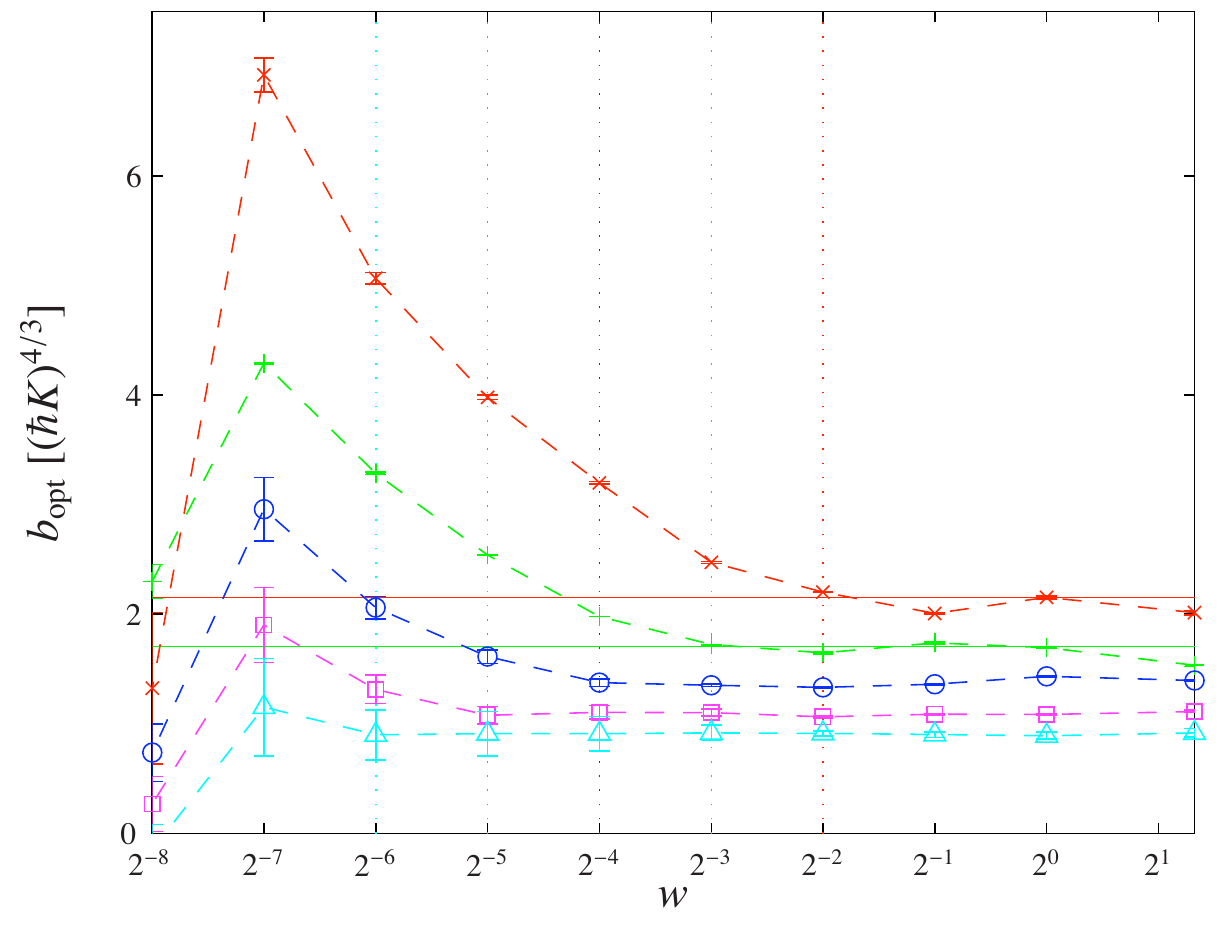}
\caption{(Color online)  The asymptote gradient $b_{\rm opt}$ of $\langle\langle \hat{p}^4 \rangle\rangle^{1/3}$.  Parameters are ${\cal N}=10000$, $T = T_T$ ($\ell=2$), $\phi_d = 0.8\pi$, and 
      ({\color{red}          $\times$}) $\Omega = 0$,
({\color{green}      $+$}) $\Omega = 1/2$,
 ({\color{blue}         $\circ$}) $\Omega = 1/4$,
 ({\color{magenta} $\square$}) $\Omega = 1/8$, and
({\color{cyan}         $\triangle$}) $\Omega = 1/16$.
The dashed lines are shown to guide the eye.  The horizontal solid lines correspond to the thermal asymptote gradients of Eqs.~(\ref{eq:c4_i_hot}) and (\ref{eq:c4_h_hot}).
The vertical dotted lines correspond to Eq.~(\ref{w_th}).}
\label{fig:gradient}
\end{figure}

We observe that for larger values of $s$ (where $\Omega=1/s$), the cumulant evolution tends to the large-$w$ thermal limit behavior more readily, i.e., for smaller values of $w$.  We have investigated this quantitatively by determining the asymptotic dependence of the fourth-order cumulant.  This was done by fitting a particular number of asymptotic points using a least-squares method according to $\langle \langle \hat{p}^4 \rangle \rangle_n ^{1/3} = bn+c$.  The most appropriate asymptote, with $b=b_{\rm opt}$, was identified by choosing the number of asymptotic points such that the standard error in the gradient $b$ was minimized.  

In Fig.\ \ref{fig:gradient} we show the optimum asymptote gradient $b_{\rm opt}$ as the width $w$ of the Gaussian distribution is varied.  We find that as $w$ increases, the asymptote gradient saturates to its thermal limit value $b_{\rm th}$.  For integer and half-integer values of $\Omega$, $b_{\rm th}$ can be determined from Eqs.~(\ref{eq:c4_i_hot}) and (\ref{eq:c4_h_hot}) to be $b_{\rm th} = (\phi_d^4 /4)^{1/3} (\hbar K)^{4/3}$ and  $b_{\rm th} = (\phi_d^4 /8)^{1/3}(\hbar K)^{4/3}$, respectively, as indicated by the solid horizontal lines in Fig.\ \ref{fig:gradient}.    

As previously indicated, the asymptote gradient saturates to the thermal value $b_{\rm th}$ more readily for higher-order fractional resonances, i.e., for larger values of $s$ (where $\Omega=1/s$).  This saturation occurs when the atomic momentum distribution initially spans a single fractional resonance feature in $\beta$, as described in Sec.\ \ref{sec:t_dep_moments}.  Quantitatively, the thermalization width $w_{\rm th}$ is defined by Eq.\ (\ref{w_th}), and as indicated by the vertical dotted lines in Fig.\ \ref{fig:gradient} it accurately predicts the temperature at which the fractional resonant dynamics saturates in the thermal limit.

\section{Discussion and Conclusions \label{sec:conclusions}}

We have presented a detailed theoretical investigation into fractional quantum resonances in the atom-optical $\delta$-kicked accelerator.  Fractional resonances occur for particular rational values of the scaled effective gravitational acceleration $\Omega$, and are characterized at zero temperature by expansion of the atom cloud in quasi-periodic bursts.  We have considered the effect of the initial atom cloud temperature on fractional quantum resonant dynamics, and have shown that the kick number at which the system dynamics clearly deviate from the zero-temperature case is identical for all fractional resonances and is inversely proportional to the initial atomic momentum width of the cloud.  However, higher-order fractional resonances are more sensitive to thermal effects and the dynamics saturates to the thermal limit behavior at lower temperatures.  

The observed resonant features are sensitively dependent on the difference between the local gravitational acceleration and the applied walking-wave acceleration.  As calibration of the phase shift between the two counterpropagating waves making up the walking wave to between parts-per-million and parts-per-billion is in principle possible \cite{Peters1999}, there may be a potential application in the sensitive atom-optical determination of the local gravitational acceleration \cite{Peters1999,Ma2004,Carusotto2005,Gustavsson2007}.  For a pulsing periodicity of equal to the Talbot time, one would for example observe resonant behaviour when the local gravitational acceleration is exactly cancelled by the acceleration of the walking wave.  

Typically in cold atom experiments the initial atomic momentum width is on the order of $w\sim 0.04$ for Bose-condensed rubidium \cite{Duffy2004b}, and $w\sim 0.008$ for Bose-condensed cesium \cite{Kraemer2004}.   At these temperatures the experimental observation of fractional resonances is accessible.  

\section*{Acknowledgements}

We thank the UK EPSRC (Grant  no.\ EP/D032970/1), and Durham University, for support.  We also thank C. S.  Adams, S. L. Cornish, M. Edwards, K. Helmerson, I. G. Hughes, M. P. A. Jones, W. D. Phillips, R. M. Potvliege, and M. G. Raizen, for useful discussions.

\begin{appendix}

\section{Factorization of the free-evolution operator\label{app:factor}}
\label{appendix:pz_factorisation}

To factorize the free-evolution component of the Floquet operator (\ref{eq:Floquet}) into position- and momentum-dependent parts, we first consider a general operator of the form
\begin{equation}
\hat{O}(\tau)=\exp\left(\rho \tau \hat{p}^2+\sigma \tau^2\hat{p}+\eta \tau^3\right)\exp(\gamma \tau\hat{z}).
\label{eq:isolated_exponential}
\end{equation}
Taking the partial derivative of $\hat{O}$ with respect to  $\tau$ gives
\begin{equation}
\begin{split}
\frac{\partial\hat{O}(\tau)}{\partial\tau}
=&
\left[
e^{\rho \tau \hat{p}^2+\sigma \tau^2 \hat{p}+\eta \tau^3}
\gamma
\hat{z} e^{-(\rho \tau \hat{p}^2+\sigma \tau^2 \hat{p}+\eta \tau^3)}
\right.\\
&+\left. \left(\rho \hat{p}^2 + 2 \sigma \tau \hat{p}+3 \eta \tau^2 \right)\right]
\hat{O}(\tau).
\label{eq:Oderivative}
\end{split}
\end{equation}
Using the general operator identity \cite{Louisell1973}
\begin{equation}
e^{\hat{A}}\hat{B}e^{-\hat{A}}= \hat{B} 
+ [\hat{A},\hat{B}] 
+ \frac{1}{2!}[\hat{A},[\hat{A},\hat{B}]]
+\cdots
\end{equation}
(truncated to the first two terms, as all further terms are zero), we find that
\begin{equation}
   e^{\rho \tau \hat{p}^2+
    \sigma \tau^2\hat{p}+
    \eta \tau ^3} \gamma \hat{z}e^{-(\rho \tau \hat{p}^2+\sigma \tau ^2 \hat{p}+\eta \tau^3)}=
    \gamma\hat{z}-i\hbar\gamma\left(2\rho \tau \hat{p}+\sigma \tau^2\right).
\end{equation}
Hence, (\ref{eq:Oderivative}) simplifies to
\begin{equation}
 \frac{\partial\hat{O}(\tau)}{\partial\tau}
 = 
\left[ \rho \hat{p}^2 +
2\tau\left(\sigma -i\hbar\rho \gamma \right) \hat{p}
+\gamma \hat{z} +\tau^2\left(3 \eta -i\hbar\sigma \gamma \right)
\right]
\hat{O}(\tau).
\label{eq:diffE1}
\end{equation}
Choosing $\sigma$ and $\eta$ such that terms proportional to $\tau$ and $\tau^{2}$ vanish, i.e., $\sigma = i\hbar \rho\gamma $, and subsequently $\eta =-\hbar^{2}\rho\gamma^{2}/3$, Eq.\ (\ref{eq:diffE1}) reduces to 
\begin{equation}
\frac{\partial\hat{O}(\tau)}{\partial\tau}
 = 
(\rho \hat{p}^2 +
+\gamma \hat{z})
\hat{O}(\tau).
\label{simple_DE}
\end{equation}
Noting from Eq.\ (\ref{eq:isolated_exponential}) that  $\hat{O}(0)=1$, Eq.\ (\ref{simple_DE}) can be integrated to give
\begin{equation}
\hat{O}(\tau)=\exp(\rho \tau \hat{p}^2 + \gamma \tau \hat{z}).
\end{equation}
Hence,
\begin{equation}
\begin{split}
\exp \left(\rho \tau \hat{p}^2+\gamma \tau\hat{z}\right)
 = &\exp\left(\rho \tau \hat{p}^2+i\hbar \rho \gamma \tau^2\hat{p}-\hbar^2 \rho \gamma^2 \tau^3/3 \right)\\
&\times \exp\left(\gamma \tau\hat{z}\right).
\end{split}
\label{eq:result}
\end{equation}
Setting $\rho = -i/2M\hbar$, $\gamma = -iM a/\hbar$, and $\tau = T$, 
yields the factorized form of the free-evolution component of the Floquet operator, as used in (\ref{first_floquet}).

\section{$\tilde{H}_{\mbox{\scriptsize stat}}$ for rational $\Omega$ \label{Tperiodicity}}

For particular values of $\Omega$, Hamiltonian\ (\ref{Hstat}) has a periodicity in addition to the delta-kicks.  This arises due to the $\Omega$-dependent phase of the standing-wave pulses, which is identical for a given remainder of  $n^{2}\Omega/2$.  In the case where $\Omega = r/s$, this occurs whenever $n^{2} r/s \bmod{2}$ takes the same value.

First we consider the case where $sr$ is even.  At the $s$th kick, i.e., where $n=s$, we have that $n^{2} r/s \bmod{2}=sr \bmod{2}=0$, and the phase of the standing-wave pulse is identical to the phase taken for $n=0$.  For $n=s+n'$, 
\begin{equation}
(s+n')^{2}\frac{r}{s} \bmod{2}
=
(2sn'+n'^{2})\frac{r}{s} \bmod{2}
=
n'^{2}\frac{r}{s} \bmod{2},
\end{equation}
and, in the frame where the cloud of atoms is (on average) stationary, there is an exact $s$-periodicity to the time dependence of the spatial phase of the standing-wave pulses.

Similarly, if $sr$ is odd, at the $2s$th kick, $n^{2} r/s \bmod{2}=4sr \bmod{2}=0$.  For  $n=2s+n'$, 
\begin{equation}
(2s+n')^{2}\frac{r}{s} \bmod{2}
=
n'^{2}\frac{r}{s} \bmod{2},
\end{equation}
and there is a $2s$-periodicity to the spatial phase of the standing-wave pulses. 

A unified expression for the temperal period $n_T$ of the phase of the standing-wave pulses (valid for even and odd $sr$), is given by $n_T=s(1 + sr\bmod{2})$.

\section{Evaluating $\nu$ for half-integer $\Omega$}
\label{appendix:find_nu}

To determine the time evolution of the momentum eigenstate (\ref{init_state}) requires knowledge of the complex amplitudes (\ref{elements}).  Theefore, we must calculate $\nu$ of Eq.\ (\ref{nu_eqn}).  In particular cases, the Gauss sum \cite{Apostol1976,Armitage2000,Bigourd2008,Gilowski2008} in $\nu$ can be evaluated analytically \cite{Halkyard2008}.  Here we evaluate $\nu$ for half-integer values of $\Omega$, i.e., $\Omega=r_2/2$.  In this case
\begin{equation}
\nu_{n,\ell}(r_2/2,\beta) = ie^{-2in\Upsilon}e^{i\pi n^2 r_2/2}\sum_{j=0}^{n-1}e^{2ij\Upsilon}e^{-i\pi j^2 r_2/2}.
\label{nu_2}
\end{equation}
For even values of $j$, $\exp(-i\pi j^2 r_2/2) = 1$, and for odd values of $j$, $\exp(-i\pi j^2 r_2/2) = \exp(-i\pi r_2/2)$.  For an even number of kicks $n=2n'$, Eq.\ (\ref{nu_2}) can be rewritten as two sums, each with the same number of elements, i.e., 
\begin{equation}
\begin{split}
\nu_{n=2n',\ell}(r_2/2,\beta)  = &  e^{-4in'\Upsilon}e^{2i\pi n'^2 r_2}
\left[\sum_{j_1=0}^{n'-1}e^{4ij_1\Upsilon} \right.\\
&  \left. +e^{-i\pi r_2/2}\sum_{j_2=0}^{n'-1}e^{2i(2j_2+1)\Upsilon}\right].
\end{split}
\end{equation}
Simplifying, we find that
\begin{equation}
\nu_{n=2n',\ell}(r_2/2,\beta) = ie^{-4in'\Upsilon}\left[1+e^{i\left(2\Upsilon-\pi r_2 / 2\right)}\right]
\sum_{j=0}^{n'-1}e^{4ij\Upsilon}.
\end{equation}
Evaluating the geometric sum, and simplifying further, yields
\begin{equation}
\begin{split}
\nu_{n=2n',\ell}(r_2/2,\beta) = & 2 i e^{-i\left[\Upsilon(2n'+1) + \pi r_2/4\right]}
\\&\times \cos\left(\Upsilon - \pi r_2/4\right)\frac{\sin(2n'\Upsilon)}{\sin(2\Upsilon)}.
\end{split}
\end{equation}
Resubstituting for $n=2n'$ gives
\begin{equation}
\nu_{n,\ell}(r_2/2,\beta) = 2 i e^{-i\left[\Upsilon(n+1) + \pi r_2/4\right]}\cos\left(\Upsilon - \pi r_2/4\right)\frac{\sin(n\Upsilon)}{\sin(2\Upsilon)},
\end{equation}
which applies for even $n$ only.

\section{Momentum moments for half-integer $\Omega$}
\label{appendix:moments}

At finite temperature, the $q$th-order momentum moment is given by Eq.\ (\ref{eq:mom_mom_dist}).  Inserting the matrix elements from Eq.\ (\ref{elements}), using that $\omega = \phi_d |\nu|$ [see Eq.\ (\ref{Eq:omega_nu})], and returning to the atomic momentum notation $p=\hbar K(k+\beta)$, we find that
\begin{equation}
\langle \hat{p}^q \rangle_n
 = \int dp
\sum_{j=-\infty}^{\infty}
J_{j}^2\left(\phi_d |\nu|\right)
D(p) 
 \left(\hbar Kj+p\right)^{q}.
\label{eq:pq_definition}
\end{equation}

Here we consider only the second- and fourth-order momentum moments in detail. The second-order momentum moment is given by
\begin{equation}
\langle \hat{p}^2 \rangle_n
 = \int dp
\sum_{j=-\infty}^{\infty}
J_{j}^2\left(\phi_d |\nu|\right)
D(p) 
 \left[(\hbar K)^2j^2+2\hbar Kjp+p^2\right].
\label{eq:p2_definition}
\end{equation}
Using the Bessel function identities $\sum_{j=-\infty}^{\infty} j^2 J_j^2(\eta) = \eta^2/2$,
$\sum_{j=-\infty}^{\infty} j J_j^2(\eta)=0$, and $\sum_{j=-\infty}^{\infty} J_j^2(\eta)=1$ \cite{Abramowitz1964, Saunders2007}, Eq.\ (\ref{eq:p2_definition}) becomes
\begin{equation}
\langle \hat{p}^2\rangle_{n}
=  \int  dp D(p) p^2 
 + (\hbar K)^2\frac{\phi_d^2}{2} \int dp D(p) |\nu|^2.
\label{eq:p2_terms}
\end{equation}

The fourth-order momentum moment is given by
\begin{equation}
\begin{split}
\langle \hat{p}^4 \rangle_{n}
 = & \int dp
\sum_{j=-\infty}^{\infty}
J_{j}^2\left(\phi_d |\nu|\right) 
D(p) \left[(\hbar K)^4j^{4} \right.
\\
&   \left. + 4(\hbar K)^3j^{3}p + 6(\hbar K)^2j^{2}p^{2} + 4\hbar Kjp^{3} + p^4 \right].
\label{eq:p4_definition}
\end{split}
\end{equation}
Using the Bessel function identities above, as well as $\sum_{j=-\infty}^{\infty}j^3 J_j^2(\eta)=0$ and $\sum_{j=-\infty}^{\infty} j^4 J_j^2(\eta) = 3\eta^4/8+\eta^2/2$ \cite{Halkyard2008}, Eq.\ (\ref{eq:p4_definition}) becomes
\begin{equation}
\begin{split}
\langle\hat{p}^4 \rangle_{n} 
= & \int p^4 D(p) dp + (\hbar K)^4\frac{\phi_d^2}{2} \int  dpD(p)|\nu|^2 
\\
&  + (\hbar K)^2 3\phi_d^2 \int dpD(p) p^2 |\nu|^2 \\
&  + (\hbar K)^4 \frac{3\phi_d^4}{8} \int dpD(p)|\nu|^4.
\end{split}
\label{eq:p4_terms}
\end{equation}

For half-integer values of $\Omega$, the second-order momentum moment evolution for a Gaussian distribution is given by Eq.\ (\ref{eq:p2_terms}) with $\nu$ of Eq.\ (\ref{half_int_nu}).  Using Eqs.\ (\ref{eq:p2_TERM1}) and (\ref{eq:p2_TERM2}), we find that
\begin{equation}
\frac{\langle \hat{p}^{2} \rangle_{n}}{(\hbar K)^2}  =
w^2 + \frac{\phi_d^2}{2} n
 + \phi_d^2 \sum_{m=1}^{n/2-1}  (n-2m)
e^{-8m^2 \pi^2\ell^2 w^2}.
\label{eq:p2_half}
\end{equation}
In the limit $w \gg 1/2\sqrt{2}\pi \ell$, the decaying terms become negligible and we can write
\begin{equation}
\frac{\langle \hat{p}^{2} \rangle_{n}}{(\hbar K)^2} =
w^2 + \frac{\phi_d^2}{2} n.
\label{eq:p2_half_2}
\end{equation}
Note that Eqs.\ (\ref{eq:p2_half}) and (\ref{eq:p2_half_2}) only apply for even $n$.

The fourth-order momentum moment for a Gaussian distribution is given by Eq.\ (\ref{eq:p4_terms}).  Using Eqs.\ (\ref{eq:p4_TERM1}), (\ref{eq:p2_TERM2}), (\ref{eq:p4_TERM4}), and (\ref{eq:p4_TERM2}), we find that
\begin{equation}
\begin{split}
\frac{\langle \hat{p}^{4} \rangle_{n}}{(\hbar K)^4} 
 =& 3w^4 +\frac{\phi_d^2}{2}n+ 3w^2\phi_d^2n + \frac{\phi_d^4}{16}n(n^2+2) \left(3- e^{-8\pi^2 w^2 \ell^2}\right) \\
 & 
+ 6\phi_d^2 w^2\sum_{m=1}^{n/2-1} (n - 2m)(1-4\pi^2m^2\ell^2w^2)e^{-8\pi^2 m^2 \ell^2 w^2} \\
&+ \frac{3\phi_d^4}{8} \sum_{m=1}^{n/2-1} \left[ m^3-n m^2-m+\frac{1}{6}n(n^2+2) \right] \\
&\times \left[ 6e^{-8\pi^2m^2\ell^2w^2} -e^{-8\pi^2  (m+1)^2 \ell^2w^2} - e^{-8\pi^2  (m-1)^2\ell^2w^2} \right] \\
&- \frac{3\phi_d^4}{8}
\sum_{m=n/2}^{n-2} \Biggl[ \frac{m^3}{3}-n m^2 +\frac{1}{3}(3n^2-1)m
+\frac{n}{3}(1-n^2) \Biggr] \\
&\times
\left[ 6e^{-8\pi^2m^2\ell^2w^2} -e^{-8\pi^2 (m+1)^2 \ell^2 w^2} -e^{-8\pi^2 (m-1)^2\ell^2 w^2} \right] .
\end{split}
\end{equation}
In the limit $w \gg 1/2\sqrt{2}\pi \ell,$ we can write
\begin{equation}
\frac{\langle \hat{p}^{4} \rangle_{n}}{(\hbar K)^4} 
 = 3w^4 +\frac{\phi_d^2}{2}n+ 3\phi_d^2w^2n + \frac{\phi_d^4}{8}  n(n^2+5).\label{eq:p4_half}
\end{equation}
Again, we emphasize that Eqs.\ (\ref{eq:p2_half_2}) and (\ref{eq:p4_half}) only apply for even $n$.

\section{Integrals over $D(p)$}

\subsection{Initial momentum moments}

At finite temperature, the initial $q$th-order momentum moment is given by Eq.\ (\ref{eq:pq_definition}) with $\nu=0$, i.e.,
\begin{equation}
\langle\hat{p}^q \rangle_{0}
= \int  dpD(p) p^q.
\end{equation}
In the case where the atomic momenta are distributed according to the Gaussian (\ref{eq:Gaussian}), the initial second-order momentum moment is
\begin{equation}
\langle \hat{p}^2 \rangle_{0}
 = (\hbar K)^2w^2,
\label{eq:p2_TERM1}
\end{equation}
and the initial fourth-order momentum moment is
\begin{equation}
\langle \hat{p}^4 \rangle_{0}
 = 3(\hbar K)^4w^4.
\label{eq:p4_TERM1}
\end{equation}

\subsection{Integrals involving $\nu$}

\subsubsection{Evaluating $\int  dpD(p) |\nu|^2 $ \label{sec:int_nu_D}}

When $\Omega$ takes half-integer values, i.e., $\Omega = r_2/2$ with $r_2$ odd, $\nu$ is given by Eq.\ (\ref{half_int_nu}) for even kick numbers $n=2n'$.  Hence,
\begin{equation}
\int dp D(p) |\nu|^2 = 
4\int  dp D(p) \frac{\sin^2 \left(2n' \Upsilon\right)}{\sin^2 \left(2\Upsilon\right)} \cos^2\left(\Upsilon - \pi r_2/4\right).
\label{eq:nu2}
\end{equation}
Using the expansion \cite{Halkyard2008}
\begin{equation}
\frac{\sin^2(nx)}{\sin^2(x)} =  n + 2\sum_{m=1}^{n-1} (n-m) \cos(2mx),
\label{eq:sin2_identity}
\end{equation}
$\cos(4m\Upsilon) =\cos(4m\pi \ell p/\hbar K),$ and 
\begin{equation}
\cos^2\left( \Upsilon - \frac{\pi r_2}{4} \right) = \frac{1}{2}+ \frac{1}{2}(-1)^{\ell + (r_2-1)/2}\sin\left(\frac{2\pi\ell p}{\hbar K} \right),
\label{eq:cos2}
\end{equation}
equation\ (\ref{eq:nu2}) becomes
\begin{equation}
\begin{split}
\int dp D(p)  |\nu|^2  = & 2n' +4\sum_{m=1}^{n'-1} (n'-m)\int dp D(p) \cos\left( \frac{4m\pi \ell p}{\hbar K}\right) \\
& + 2(-1)^{\ell+(r_2-1)/2} \int dp D(p) \sin \left(\frac{2\pi\ell p}{\hbar K} \right) 
\\ & \times
\left[ n' +2 \sum_{m=1}^{n'-1} (n'-m)\cos\left( \frac{4m\pi \ell p}{\hbar K}\right)\right],
\label{eq:nu2b}
\end{split}
\end{equation}
where we have used the normalization of $D(p)$ to evaluate the first term.  In the case where the atomic momenta are distributed according to the Gaussian\ (\ref{eq:Gaussian}), the first integral on the right-hand side of Eq.\ (\ref{eq:nu2b}) can be evaluated using 
\begin{equation}
\int dp\cos (\lambda p) e^{-p^2/\alpha^2}  = \sqrt{\pi} \alpha e^{-\lambda^2 \alpha^2/4}.
\label{cos_Gauss_int}
\end{equation}
The remaining integrals vanish due to the even parity of $D(p)$.  Therefore, we find that
\begin{equation}
\int dp D(p)  |\nu|^2 = 2n' +4\sum_{m=1}^{n'-1} (n'-m)   e^{-8m^2 \pi^2\ell^2 w^2}.
\end{equation} 
Substituting $n=2n'$ yields
\begin{equation}
\int dp D(p) |\nu|^2 = n +2\sum_{m=1}^{n/2-1} (n-2m)   e^{-8m^2 \pi^2\ell^2 w^2},
\label{eq:p2_TERM2}
\end{equation} 
where $n$ is even.

\subsubsection{Evaluating $\int dp D(p)  p^{2} |\nu|^2 $}

For $\Omega=r_2/2$ and $n=2n'$,
\begin{equation}
\int dp D(p) p^{2} |\nu|^2 = 4  \int dp D(p) p^{2}\frac{\sin ^2 (2n'\Upsilon)}{\sin^2 (2\Upsilon)} \cos^2(\Upsilon-\pi r_2/4).
\end{equation}
Following the treatment in App.\ \ref{sec:int_nu_D}, and using (\ref{eq:p2_TERM1}) and the integral
\begin{equation}
\int
p^2 \cos(\lambda p) e^{-p^2/\alpha^2} dp
=\frac{1}{2}\sqrt{\pi} \alpha^3 \left(1-\frac{\alpha^2\lambda^2}{2} \right) e^{-\lambda^2\alpha^2/4},
\label{eq:integral_p2_cos_exp}
\end{equation}
we find that for Gaussian $D(p)$ [see Eq.\ (\ref{eq:Gaussian})],
\begin{equation}
\begin{split}
\int dp D(p) p^{2} |\nu|^2   = 
& (\hbar K)^2 w^2 \Biggl[ 2n' 
+ 4 \sum_{m=1}^{n'-1} (n' - m) \\
& \times (1-4m^2\pi^2\ell^2w^2)e^{-8m^2\pi^2 \ell^2 w^2} \Biggr].
\end{split}
\end{equation}
Substituting $n=2n'$ yields
\begin{equation}
\begin{split}
\int dp D(p) p^{2} |\nu|^2   = 
& (\hbar K)^2 w^2 \Biggr[ n 
+ 2 \sum_{m=1}^{n/2-1} (n -2 m) \\
 & \times (1-4m^2\pi^2\ell^2w^2)e^{-8m^2\pi^2 \ell^2 w^2} \Biggr].
 \end{split}
\label{eq:p4_TERM4}
\end{equation}

\subsubsection{Evaluating $\int  dp D(p) |\nu|^4$}
 
For $\Omega=r_2/2$ and $n=2n'$,
\begin{equation}
\int  dp D(p) |\nu|^4= 16\int  dp D(p) \frac{\sin^4(2n'\Upsilon)}{\sin^4(2\Upsilon)} \cos^4(\Upsilon-\pi r_2/4).
\end{equation}
Using the expansion \cite{Halkyard2008}
\begin{equation}
\begin{split}
\frac{\sin^4(nx)}{\sin^4(x)} =
& \frac{n}{3}(2n^2+1) + \sum_{m=1}^{n-1} \cos(2mx)
\\&\times \left[ m^3-2n m^2-m 
+ \frac{2n}{3}(2n^2+1) \right] 
\\&
-  \sum_{m=n}^{2n-2} \cos(2mx)
\\&\times 
\left[ \frac{m^3}{3}-2n m^2 
+\left(4n^2-\frac{1}{3}\right)m+\frac{2n}{3}(1-4n^2)\right],
\label{eq:sin4}
\end{split}
\end{equation}
together with
\begin{equation}
\begin{split}
\cos^4\left( \Upsilon - \frac{\pi r_2}{4} \right)  = 
& \frac{3}{8}- \frac{1}{8}\cos\left(\frac{4\pi \ell p}{\hbar K}\right) \\
&  + \frac{1}{2}(-1)^{\ell+(r_2-1)/2} \sin\left( \frac{2\pi\ell p}{\hbar K}\right ),
\label{eq:cos4}
\end{split}
\end{equation}
and following the treatment in App.\ \ref{sec:int_nu_D}, we find that
 \begin{equation}
 \begin{split}
\int dp D(p) |\nu|^4
 =&   \frac{2}{3} n'(2n'^2+1) \left[3- e^{-8\pi^2  \ell^2 w^2}\right]
  \\
 &+  \sum_{m=1}^{n'-1} \left[ m^3-2n' m^2-m+\frac{2n'}{3}(2n'^2+1) \right] \\
& \times  \Bigl[ 6 e^{-8\pi^2m^2\ell^2w^2} -e^{-8\pi^2 (m+1)^2 \ell^2w^2}
\\&- e^{-8\pi^2 (m-1)^2 \ell^2 w^2}  \Bigr] \\
 & - \sum_{m=n'}^{2n'-2} \left[ \frac{m^3}{3}-2n' m^2 +\frac{1}{3}(12n'^2-1)m \right. \\
& \left. + \frac{2n'}{3}(1-4n'^2) \right] \left[ 6 e^{-8\pi^2m^2\ell^2w^2} \right. \\
& \left. -e^{-8\pi^2 (m+1)^2 \ell^2w^2}- e^{-8\pi^2 (m-1)^2 \ell^2 w^2}  \right].
\end{split}
\end{equation}
Substituting $n=2n'$ then yields
 \begin{equation}
 \begin{split}
\int dp D(p) |\nu|^4
 =&   \frac{1}{6} n(n^2+2) \left[3- e^{-8\pi^2  \ell^2 w^2}\right]
  \\
& +  \sum_{m=1}^{n/2-1} \left[ m^3-n m^2-m+\frac{1}{6}n (n^2+2) \right] \\
& \times  \Bigl[ 6 e^{-8\pi^2m^2\ell^2w^2} -e^{-8\pi^2 (m+1)^2 \ell^2w^2}
\\ & - e^{-8\pi^2 (m-1)^2 \ell^2 w^2}  \Bigr] \\
&  - \sum_{m=n/2}^{n-2}  \Biggl[\frac{m^3}{3}-n m^2 +\frac{1}{3}(3n^2-1)m \\
& +\frac{n}{3}(1-n^2) \Biggr] \Bigl[ 6 e^{-8\pi^2m^2\ell^2w^2}\\
& -e^{-8\pi^2 (m+1)^2 \ell^2w^2}- e^{-8\pi^2 (m-1)^2 \ell^2 w^2}  \Bigr].
\label{eq:p4_TERM2}
\end{split}
\end{equation}

\end{appendix}



\begin{thebibliography}{74}
\expandafter\ifx\csname natexlab\endcsname\relax\def\natexlab#1{#1}\fi
\expandafter\ifx\csname bibnamefont\endcsname\relax
  \def\bibnamefont#1{#1}\fi
\expandafter\ifx\csname bibfnamefont\endcsname\relax
  \def\bibfnamefont#1{#1}\fi
\expandafter\ifx\csname citenamefont\endcsname\relax
  \def\citenamefont#1{#1}\fi
\expandafter\ifx\csname url\endcsname\relax
  \def\url#1{\texttt{#1}}\fi
\expandafter\ifx\csname urlprefix\endcsname\relax\def\urlprefix{URL }\fi
\providecommand{\bibinfo}[2]{#2}
\providecommand{\eprint}[2][]{\url{#2}}

\bibitem[{\citenamefont{{W.~H.~Oskay} et~al.}(2000)\citenamefont{{W.~H.~Oskay},
  {D.~A.~Steck}, {V.~Milner}, {B.~G.~Klappauf}, and
  {M.~G.~Raizen}}}]{Oskay2000}
\bibinfo{author}{\bibnamefont{{W.~H.~Oskay}}},
  \bibinfo{author}{\bibnamefont{{D.~A.~Steck}}},
  \bibinfo{author}{\bibnamefont{{V.~Milner}}},
  \bibinfo{author}{\bibnamefont{{B.~G.~Klappauf}}}, \bibnamefont{and}
  \bibinfo{author}{\bibnamefont{{M.~G.~Raizen}}}, \bibinfo{journal}{Opt.~Comm.}
  \textbf{\bibinfo{volume}{179}}, \bibinfo{pages}{137} (\bibinfo{year}{2000}).

\bibitem[{\citenamefont{{M.~Sadgrove} et~al.}(2004)\citenamefont{{M.~Sadgrove},
  {A.~Hilliard}, {T.~Mullins}, {S.~Parkins}, and
  {R.~Leonhardt}}}]{Sadgrove2004}
\bibinfo{author}{\bibnamefont{{M.~Sadgrove}}},
  \bibinfo{author}{\bibnamefont{{A.~Hilliard}}},
  \bibinfo{author}{\bibnamefont{{T.~Mullins}}},
  \bibinfo{author}{\bibnamefont{{S.~Parkins}}}, \bibnamefont{and}
  \bibinfo{author}{\bibnamefont{{R.~Leonhardt}}},
  \bibinfo{journal}{Phys.~Rev.~E} \textbf{\bibinfo{volume}{70}},
  \bibinfo{pages}{036217} (\bibinfo{year}{2004}).

\bibitem[{\citenamefont{{C.~F.~Bharucha}
  et~al.}(1999)\citenamefont{{C.~F.~Bharucha}, {J.~C.~Robinson}, {F.~L.~Moore},
  {B.~Sundaram}, {Q.~Niu}, and {M.~G.~Raizen}}}]{Bharucha1999}
\bibinfo{author}{\bibnamefont{{C.~F.~Bharucha}}},
  \bibinfo{author}{\bibnamefont{{J.~C.~Robinson}}},
  \bibinfo{author}{\bibnamefont{{F.~L.~Moore}}},
  \bibinfo{author}{\bibnamefont{{B.~Sundaram}}},
  \bibinfo{author}{\bibnamefont{{Q.~Niu}}}, \bibnamefont{and}
  \bibinfo{author}{\bibnamefont{{M.~G.~Raizen}}},
  \bibinfo{journal}{Phys.~Rev.~E} \textbf{\bibinfo{volume}{60}},
  \bibinfo{pages}{3881} (\bibinfo{year}{1999}).

\bibitem[{\citenamefont{{M.~B.~d'Arcy}
  et~al.}(2001{\natexlab{a}})\citenamefont{{M.~B.~d'Arcy}, {R.~M.~Godun},
  {M.~K.~Oberthaler}, {G~ S.~Summy}, {K.~Burnett}, and
  {S.~A.~Gardiner}}}]{dArcy2001a}
\bibinfo{author}{\bibnamefont{{M.~B.~d'Arcy}}},
  \bibinfo{author}{\bibnamefont{{R.~M.~Godun}}},
  \bibinfo{author}{\bibnamefont{{M.~K.~Oberthaler}}},
  \bibinfo{author}{\bibnamefont{{G~ S.~Summy}}},
  \bibinfo{author}{\bibnamefont{{K.~Burnett}}}, \bibnamefont{and}
  \bibinfo{author}{\bibnamefont{{S.~A.~Gardiner}}},
  \bibinfo{journal}{Phys.~Rev.~E} \textbf{\bibinfo{volume}{64}},
  \bibinfo{pages}{056233} (\bibinfo{year}{2001}{\natexlab{a}}).

\bibitem[{\citenamefont{{M.~K.~Oberthaler}
  et~al.}(1999)\citenamefont{{M.~K.~Oberthaler}, {R.~M.~Godun}, {M.~B.~d'Arcy},
  {G.~S.~Summy}, and {K.~Burnett}}}]{Oberthaler1999}
\bibinfo{author}{\bibnamefont{{M.~K.~Oberthaler}}},
  \bibinfo{author}{\bibnamefont{{R.~M.~Godun}}},
  \bibinfo{author}{\bibnamefont{{M.~B.~d'Arcy}}},
  \bibinfo{author}{\bibnamefont{{G.~S.~Summy}}}, \bibnamefont{and}
  \bibinfo{author}{\bibnamefont{{K.~Burnett}}},
  \bibinfo{journal}{Phys.~Rev.~Lett.} \textbf{\bibinfo{volume}{83}},
  \bibinfo{pages}{4447} (\bibinfo{year}{1999}).

\bibitem[{\citenamefont{{R.~M.~Godun} et~al.}(2000)\citenamefont{{R.~M.~Godun},
  {M.~B.~d'Arcy}, {M.~K.~Oberthaler}, {G.~S.~Summy}, and
  {K.~Burnett}}}]{Godun2000}
\bibinfo{author}{\bibnamefont{{R.~M.~Godun}}},
  \bibinfo{author}{\bibnamefont{{M.~B.~d'Arcy}}},
  \bibinfo{author}{\bibnamefont{{M.~K.~Oberthaler}}},
  \bibinfo{author}{\bibnamefont{{G.~S.~Summy}}}, \bibnamefont{and}
  \bibinfo{author}{\bibnamefont{{K.~Burnett}}}, \bibinfo{journal}{Phys.~Rev.~A}
  \textbf{\bibinfo{volume}{62}}, \bibinfo{pages}{013411}
  (\bibinfo{year}{2000}).

\bibitem[{\citenamefont{{S.~Schlunk}
  et~al.}(2003{\natexlab{a}})\citenamefont{{S.~Schlunk}, {M.~B.~d'Arcy},
  {S.~A.~Gardiner}, {D.~Cassettari}, {R.~M.~Godun}, and
  {G.~S.~Summy}}}]{Schlunk2003a}
\bibinfo{author}{\bibnamefont{{S.~Schlunk}}},
  \bibinfo{author}{\bibnamefont{{M.~B.~d'Arcy}}},
  \bibinfo{author}{\bibnamefont{{S.~A.~Gardiner}}},
  \bibinfo{author}{\bibnamefont{{D.~Cassettari}}},
  \bibinfo{author}{\bibnamefont{{R.~M.~Godun}}}, \bibnamefont{and}
  \bibinfo{author}{\bibnamefont{{G.~S.~Summy}}},
  \bibinfo{journal}{Phys.~Rev.~Lett.} \textbf{\bibinfo{volume}{90}},
  \bibinfo{pages}{054101} (\bibinfo{year}{2003}{\natexlab{a}}).

\bibitem[{\citenamefont{{S.~Schlunk}
  et~al.}(2003{\natexlab{b}})\citenamefont{{S.~Schlunk}, {M.~B.~d'Arcy},
  {S.~A.~Gardiner}, and {G.~S.~Summy}}}]{Schlunk2003b}
\bibinfo{author}{\bibnamefont{{S.~Schlunk}}},
  \bibinfo{author}{\bibnamefont{{M.~B.~d'Arcy}}},
  \bibinfo{author}{\bibnamefont{{S.~A.~Gardiner}}}, \bibnamefont{and}
  \bibinfo{author}{\bibnamefont{{G.~S.~Summy}}},
  \bibinfo{journal}{Phys.~Rev.~Lett.} \textbf{\bibinfo{volume}{90}},
  \bibinfo{pages}{124102} (\bibinfo{year}{2003}{\natexlab{b}}).

\bibitem[{\citenamefont{{Z.-Y.~Ma} et~al.}(2004)\citenamefont{{Z.-Y.~Ma},
  {M.~B.~d'Arcy}, and {S.~A.~Gardiner}}}]{Ma2004}
\bibinfo{author}{\bibnamefont{{Z.-Y.~Ma}}},
  \bibinfo{author}{\bibnamefont{{M.~B.~d'Arcy}}}, \bibnamefont{and}
  \bibinfo{author}{\bibnamefont{{S.~A.~Gardiner}}},
  \bibinfo{journal}{Phys.~Rev.~Lett.} \textbf{\bibinfo{volume}{93}},
  \bibinfo{pages}{164101} (\bibinfo{year}{2004}).

\bibitem[{\citenamefont{{A.~Buchleitner}
  et~al.}(2006)\citenamefont{{A.~Buchleitner}, {M.~B.~d'Arcy}, {S.~Fishman},
  {S.~A.~Gardiner}, {I.~Guarneri}, {Z.-Y.~Ma}, {L.~Rebuzzini}, and
  {G.~S.~Summy}}}]{Buchleitner2006}
\bibinfo{author}{\bibnamefont{{A.~Buchleitner}}},
  \bibinfo{author}{\bibnamefont{{M.~B.~d'Arcy}}},
  \bibinfo{author}{\bibnamefont{{S.~Fishman}}},
  \bibinfo{author}{\bibnamefont{{S.~A.~Gardiner}}},
  \bibinfo{author}{\bibnamefont{{I.~Guarneri}}},
  \bibinfo{author}{\bibnamefont{{Z.-Y.~Ma}}},
  \bibinfo{author}{\bibnamefont{{L.~Rebuzzini}}}, \bibnamefont{and}
  \bibinfo{author}{\bibnamefont{{G.~S.~Summy}}},
  \bibinfo{journal}{Phys.~Rev.~Lett.} \textbf{\bibinfo{volume}{96}},
  \bibinfo{pages}{164101} (\bibinfo{year}{2006}).

\bibitem[{\citenamefont{{G.~Behinaein}
  et~al.}(2006)\citenamefont{{G.~Behinaein}, {V.~Ramareddy}, {P.~Ahmadi}, and
  {G.~S.~Summy}}}]{Behinaein2006}
\bibinfo{author}{\bibnamefont{{G.~Behinaein}}},
  \bibinfo{author}{\bibnamefont{{V.~Ramareddy}}},
  \bibinfo{author}{\bibnamefont{{P.~Ahmadi}}}, \bibnamefont{and}
  \bibinfo{author}{\bibnamefont{{G.~S.~Summy}}},
  \bibinfo{journal}{Phys.~Rev.~Lett.} \textbf{\bibinfo{volume}{97}},
  \bibinfo{pages}{244101} (\bibinfo{year}{2006}).

\bibitem[{\citenamefont{{I.~Guarneri} and {L.~Rebuzzini}}(2008)}]{Guarneri2008}
\bibinfo{author}{\bibnamefont{{I.~Guarneri}}} \bibnamefont{and}
  \bibinfo{author}{\bibnamefont{{L.~Rebuzzini}}},
  \bibinfo{journal}{Phys.~Rev.~Lett.} \textbf{\bibinfo{volume}{100}},
  \bibinfo{pages}{234103} (\bibinfo{year}{2008}).

\bibitem[{\citenamefont{{F.~L.~Moore} et~al.}(1995)\citenamefont{{F.~L.~Moore},
  {J.~C.~Robinson}, {C.~F.~Bharucha}, {B.~Sundaram}, and
  {M.~G.~Raizen}}}]{Moore1995}
\bibinfo{author}{\bibnamefont{{F.~L.~Moore}}},
  \bibinfo{author}{\bibnamefont{{J.~C.~Robinson}}},
  \bibinfo{author}{\bibnamefont{{C.~F.~Bharucha}}},
  \bibinfo{author}{\bibnamefont{{B.~Sundaram}}}, \bibnamefont{and}
  \bibinfo{author}{\bibnamefont{{M.~G.~Raizen}}},
  \bibinfo{journal}{Phys.~Rev.~Lett.} \textbf{\bibinfo{volume}{75}},
  \bibinfo{pages}{4598} (\bibinfo{year}{1995}).

\bibitem[{\citenamefont{{H.~Ammann} et~al.}(1998)\citenamefont{{H.~Ammann},
  {R.~Gray}, {I.~Shvarchuck}, and {N.~Christensen}}}]{Ammann1998a}
\bibinfo{author}{\bibnamefont{{H.~Ammann}}},
  \bibinfo{author}{\bibnamefont{{R.~Gray}}},
  \bibinfo{author}{\bibnamefont{{I.~Shvarchuck}}}, \bibnamefont{and}
  \bibinfo{author}{\bibnamefont{{N.~Christensen}}},
  \bibinfo{journal}{Phys.~Rev.~Lett.} \textbf{\bibinfo{volume}{80}},
  \bibinfo{pages}{4111} (\bibinfo{year}{1998}).

\bibitem[{\citenamefont{{M.~B.~d'Arcy}
  et~al.}(2001{\natexlab{b}})\citenamefont{{M.~B.~d'Arcy}, {R.~M.~Godun},
  {M.~K.~Oberthaler}, {D.~Cassettari}, and {G.~S.~Summy}}}]{dArcy2001b}
\bibinfo{author}{\bibnamefont{{M.~B.~d'Arcy}}},
  \bibinfo{author}{\bibnamefont{{R.~M.~Godun}}},
  \bibinfo{author}{\bibnamefont{{M.~K.~Oberthaler}}},
  \bibinfo{author}{\bibnamefont{{D.~Cassettari}}}, \bibnamefont{and}
  \bibinfo{author}{\bibnamefont{{G.~S.~Summy}}},
  \bibinfo{journal}{Phys.~Rev.~Lett.} \textbf{\bibinfo{volume}{87}},
  \bibinfo{pages}{074102} (\bibinfo{year}{2001}{\natexlab{b}}).

\bibitem[{\citenamefont{{F.~L.~Moore} et~al.}(1994)\citenamefont{{F.~L.~Moore},
  {J.~C.~Robinson}, {C.~Bharucha}, {P.~E.~Williams}, and
  {M.~G.~Raizen}}}]{Moore1994}
\bibinfo{author}{\bibnamefont{{F.~L.~Moore}}},
  \bibinfo{author}{\bibnamefont{{J.~C.~Robinson}}},
  \bibinfo{author}{\bibnamefont{{C.~Bharucha}}},
  \bibinfo{author}{\bibnamefont{{P.~E.~Williams}}}, \bibnamefont{and}
  \bibinfo{author}{\bibnamefont{{M.~G.~Raizen}}}, \bibinfo{journal}{Phys.~Rev.~
  Lett.} \textbf{\bibinfo{volume}{73}}, \bibinfo{pages}{2974}
  (\bibinfo{year}{1994}).

\bibitem[{\citenamefont{{B.~G.~Klappauf}
  et~al.}(1999)\citenamefont{{B.~G.~Klappauf}, {W.~H.~Oskay}, {D.~A.~Steck},
  and {M.~G.~Raizen}}}]{Klappauf1999}
\bibinfo{author}{\bibnamefont{{B.~G.~Klappauf}}},
  \bibinfo{author}{\bibnamefont{{W.~H.~Oskay}}},
  \bibinfo{author}{\bibnamefont{{D.~A.~Steck}}}, \bibnamefont{and}
  \bibinfo{author}{\bibnamefont{{M.~G.~Raizen}}}, \bibinfo{journal}{Physica~D}
  \textbf{\bibinfo{volume}{131}}, \bibinfo{pages}{78} (\bibinfo{year}{1999}).

\bibitem[{\citenamefont{{D.~A.~Steck} et~al.}(2000)\citenamefont{{D.~A.~Steck},
  {V.~Milner}, {W.~H.~Oskay}, and {M.~G.~Raizen}}}]{Steck2000}
\bibinfo{author}{\bibnamefont{{D.~A.~Steck}}},
  \bibinfo{author}{\bibnamefont{{V.~Milner}}},
  \bibinfo{author}{\bibnamefont{{W.~H.~Oskay}}}, \bibnamefont{and}
  \bibinfo{author}{\bibnamefont{{M.~G.~Raizen}}},
  \bibinfo{journal}{Phys.~Rev.~E} \textbf{\bibinfo{volume}{62}},
  \bibinfo{pages}{3461} (\bibinfo{year}{2000}).

\bibitem[{\citenamefont{{V.~Milner} et~al.}(2000)\citenamefont{{V.~Milner},
  {D.~A.~Steck}, {W.~H.~Oskay}, and {M.~G.~Raizen}}}]{Milner2000}
\bibinfo{author}{\bibnamefont{{V.~Milner}}},
  \bibinfo{author}{\bibnamefont{{D.~A.~Steck}}},
  \bibinfo{author}{\bibnamefont{{W.~H.~Oskay}}}, \bibnamefont{and}
  \bibinfo{author}{\bibnamefont{{M.~G.~Raizen}}},
  \bibinfo{journal}{Phys.~Rev.~E} \textbf{\bibinfo{volume}{61}},
  \bibinfo{pages}{7223} (\bibinfo{year}{2000}).

\bibitem[{\citenamefont{{W.~H.~Oskay} et~al.}(2003)\citenamefont{{W.~H.~Oskay},
  {D.~A.~Steck}, and {M.~G.~Raizen}}}]{Oskay2003}
\bibinfo{author}{\bibnamefont{{W.~H.~Oskay}}},
  \bibinfo{author}{\bibnamefont{{D.~A.~Steck}}}, \bibnamefont{and}
  \bibinfo{author}{\bibnamefont{{M.~G.~Raizen}}}, \bibinfo{journal}{Chaos,
  Solitons \& Fractals} \textbf{\bibinfo{volume}{16}}, \bibinfo{pages}{409}
  (\bibinfo{year}{2003}).

\bibitem[{\citenamefont{{K.~Vant} et~al.}(2000)\citenamefont{{K.~Vant},
  {G.~Ball}, and {N.~Christensen}}}]{Vant2000}
\bibinfo{author}{\bibnamefont{{K.~Vant}}},
  \bibinfo{author}{\bibnamefont{{G.~Ball}}}, \bibnamefont{and}
  \bibinfo{author}{\bibnamefont{{N.~Christensen}}},
  \bibinfo{journal}{Phys.~Rev.~E} \textbf{\bibinfo{volume}{61}},
  \bibinfo{pages}{5994} (\bibinfo{year}{2000}).

\bibitem[{\citenamefont{{A.~C.~Doherty}
  et~al.}(2000)\citenamefont{{A.~C.~Doherty}, {K.~M.~D.~Vant}, {G.~H.~Ball},
  {N.~Christensen}, and {R.~Leonhardt}}}]{Doherty2000}
\bibinfo{author}{\bibnamefont{{A.~C.~Doherty}}},
  \bibinfo{author}{\bibnamefont{{K.~M.~D.~Vant}}},
  \bibinfo{author}{\bibnamefont{{G.~H.~Ball}}},
  \bibinfo{author}{\bibnamefont{{N.~Christensen}}}, \bibnamefont{and}
  \bibinfo{author}{\bibnamefont{{R.~Leonhardt}}}, \bibinfo{journal}{J.~Opt.~B}
  \textbf{\bibinfo{volume}{2}}, \bibinfo{pages}{605} (\bibinfo{year}{2000}).

\bibitem[{\citenamefont{{J.~F.~Kanem} et~al.}(2007)\citenamefont{{J.~F.~Kanem},
  {S.~Maneshi}, {M.~Partlow}, {M.~Spanner}, and {A.~M.~Steinberg}}}]{Kanem2007}
\bibinfo{author}{\bibnamefont{{J.~F.~Kanem}}},
  \bibinfo{author}{\bibnamefont{{S.~Maneshi}}},
  \bibinfo{author}{\bibnamefont{{M.~Partlow}}},
  \bibinfo{author}{\bibnamefont{{M.~Spanner}}}, \bibnamefont{and}
  \bibinfo{author}{\bibnamefont{{A.~M.~Steinberg}}},
  \bibinfo{journal}{Phys.~Rev.~Lett.} \textbf{\bibinfo{volume}{98}},
  \bibinfo{pages}{083004} (\bibinfo{year}{2007}).

\bibitem[{\citenamefont{{M.~B.~d'Arcy}
  et~al.}(2003)\citenamefont{{M.~B.~d'Arcy}, {R.~M.~Godun}, {D.~Cassettari},
  and {G~ S.~Summy}}}]{dArcy2003}
\bibinfo{author}{\bibnamefont{{M.~B.~d'Arcy}}},
  \bibinfo{author}{\bibnamefont{{R.~M.~Godun}}},
  \bibinfo{author}{\bibnamefont{{D.~Cassettari}}}, \bibnamefont{and}
  \bibinfo{author}{\bibnamefont{{G~ S.~Summy}}},
  \bibinfo{journal}{Phys.~Rev.~A} \textbf{\bibinfo{volume}{67}},
  \bibinfo{pages}{023605} (\bibinfo{year}{2003}).

\bibitem[{\citenamefont{{G.~J.~Duffy}
  et~al.}(2004{\natexlab{a}})\citenamefont{{G.~J.~Duffy}, {A.~S.~Mellish},
  {K.~J.~Challis}, and {A.~C.~Wilson}}}]{Duffy2004b}
\bibinfo{author}{\bibnamefont{{G.~J.~Duffy}}},
  \bibinfo{author}{\bibnamefont{{A.~S.~Mellish}}},
  \bibinfo{author}{\bibnamefont{{K.~J.~Challis}}}, \bibnamefont{and}
  \bibinfo{author}{\bibnamefont{{A.~C.~Wilson}}},
  \bibinfo{journal}{Phys.~Rev.~A} \textbf{\bibinfo{volume}{70}},
  \bibinfo{pages}{041602(R)} (\bibinfo{year}{2004}{\natexlab{a}}).

\bibitem[{\citenamefont{{C.~Ryu} et~al.}(2006)\citenamefont{{C.~Ryu},
  {M.~F.~Andersen}, {A.~Vaziri}, {M.~B.~d'Arcy}, {J.~M.~Grossman},
  {K.~Helmerson}, and {W.~D.~Phillips}}}]{Ryu2006}
\bibinfo{author}{\bibnamefont{{C.~Ryu}}},
  \bibinfo{author}{\bibnamefont{{M.~F.~Andersen}}},
  \bibinfo{author}{\bibnamefont{{A.~Vaziri}}},
  \bibinfo{author}{\bibnamefont{{M.~B.~d'Arcy}}},
  \bibinfo{author}{\bibnamefont{{J.~M.~Grossman}}},
  \bibinfo{author}{\bibnamefont{{K.~Helmerson}}}, \bibnamefont{and}
  \bibinfo{author}{\bibnamefont{{W.~D.~Phillips}}},
  \bibinfo{journal}{Phys.~Rev.~Lett.} \textbf{\bibinfo{volume}{96}},
  \bibinfo{pages}{160403} (\bibinfo{year}{2006}).

\bibitem[{\citenamefont{{P.~Szriftgiser}
  et~al.}(2002)\citenamefont{{P.~Szriftgiser}, {J.~Ringot}, {D.~Delande}, and
  {J.~C.~Garreau}}}]{Szriftgiser2002}
\bibinfo{author}{\bibnamefont{{P.~Szriftgiser}}},
  \bibinfo{author}{\bibnamefont{{J.~Ringot}}},
  \bibinfo{author}{\bibnamefont{{D.~Delande}}}, \bibnamefont{and}
  \bibinfo{author}{\bibnamefont{{J.~C.~Garreau}}},
  \bibinfo{journal}{Phys.~Rev.~Lett.} \textbf{\bibinfo{volume}{89}},
  \bibinfo{pages}{224101} (\bibinfo{year}{2002}).

\bibitem[{\citenamefont{{H.~Ammann} and {N.~Christensen}}(1998)}]{Ammann1998b}
\bibinfo{author}{\bibnamefont{{H.~Ammann}}} \bibnamefont{and}
  \bibinfo{author}{\bibnamefont{{N.~Christensen}}},
  \bibinfo{journal}{Phys.~Rev.~E} \textbf{\bibinfo{volume}{57}},
  \bibinfo{pages}{354} (\bibinfo{year}{1998}).

\bibitem[{\citenamefont{{K.~Vant} et~al.}(1999)\citenamefont{{K.~Vant},
  {G.~Ball}, {H.~Ammann}, and {N.~Christensen}}}]{Vant1999}
\bibinfo{author}{\bibnamefont{{K.~Vant}}},
  \bibinfo{author}{\bibnamefont{{G.~Ball}}},
  \bibinfo{author}{\bibnamefont{{H.~Ammann}}}, \bibnamefont{and}
  \bibinfo{author}{\bibnamefont{{N.~Christensen}}},
  \bibinfo{journal}{Phys.~Rev.~E} \textbf{\bibinfo{volume}{59}},
  \bibinfo{pages}{2846} (\bibinfo{year}{1999}).

\bibitem[{\citenamefont{{M.~E.~K.~Williams}
  et~al.}(2004)\citenamefont{{M.~E.~K.~Williams}, {M.~P.~Sadgrove},
  {A.~J.~Daley}, {R.~N.~C.~Gray}, {S.~M.~Tan}, {A.~S.~Parkins},
  {N.~Christensen}, and {R.~Leonhardt}}}]{Williams2004}
\bibinfo{author}{\bibnamefont{{M.~E.~K.~Williams}}},
  \bibinfo{author}{\bibnamefont{{M.~P.~Sadgrove}}},
  \bibinfo{author}{\bibnamefont{{A.~J.~Daley}}},
  \bibinfo{author}{\bibnamefont{{R.~N.~C.~Gray}}},
  \bibinfo{author}{\bibnamefont{{S.~M.~Tan}}},
  \bibinfo{author}{\bibnamefont{{A.~S.~Parkins}}},
  \bibinfo{author}{\bibnamefont{{N.~Christensen}}}, \bibnamefont{and}
  \bibinfo{author}{\bibnamefont{{R.~Leonhardt}}}, \bibinfo{journal}{J.~Opt.~B}
  \textbf{\bibinfo{volume}{6}}, \bibinfo{pages}{28} (\bibinfo{year}{2004}).

\bibitem[{\citenamefont{{G.~J.~Duffy}
  et~al.}(2004{\natexlab{b}})\citenamefont{{G.~J.~Duffy}, {S.~Parkins},
  {T.~M\"{u}ller}, {M.~Sadgrove}, {R.~Leonhardt}, and
  {A.~C.~Wilson}}}]{Duffy2004a}
\bibinfo{author}{\bibnamefont{{G.~J.~Duffy}}},
  \bibinfo{author}{\bibnamefont{{S.~Parkins}}},
  \bibinfo{author}{\bibnamefont{{T.~M\"{u}ller}}},
  \bibinfo{author}{\bibnamefont{{M.~Sadgrove}}},
  \bibinfo{author}{\bibnamefont{{R.~Leonhardt}}}, \bibnamefont{and}
  \bibinfo{author}{\bibnamefont{{A.~C.~Wilson}}},
  \bibinfo{journal}{Phys.~Rev.~E} \textbf{\bibinfo{volume}{70}},
  \bibinfo{pages}{056206} (\bibinfo{year}{2004}{\natexlab{b}}).

\bibitem[{\citenamefont{{A.~Tonyushkin} et~al.}()\citenamefont{{A.~Tonyushkin},
  {S.~Wu}, and {M.~Prentiss}}}]{Tonyushkin2008}
\bibinfo{author}{\bibnamefont{{A.~Tonyushkin}}},
  \bibinfo{author}{\bibnamefont{{S.~Wu}}}, \bibnamefont{and}
  \bibinfo{author}{\bibnamefont{{M.~Prentiss}}}, \eprint{arXiv:0803.4153v1}.

\bibitem[{\citenamefont{{S.~Wu} et~al.}()\citenamefont{{S.~Wu},
  {A.~Tonyushkin}, and {M.~Prentiss}}}]{Wu2008}
\bibinfo{author}{\bibnamefont{{S.~Wu}}},
  \bibinfo{author}{\bibnamefont{{A.~Tonyushkin}}}, \bibnamefont{and}
  \bibinfo{author}{\bibnamefont{{M.~Prentiss}}}, \eprint{arXiv:0801.0475}.

\bibitem[{\citenamefont{{P.~H.~Jones} et~al.}(2004)\citenamefont{{P.~H.~Jones},
  {M.~M.~Stocklin}, {G.~Hur}, and {T.~S.~Monteiro}}}]{Jones2004}
\bibinfo{author}{\bibnamefont{{P.~H.~Jones}}},
  \bibinfo{author}{\bibnamefont{{M.~M.~Stocklin}}},
  \bibinfo{author}{\bibnamefont{{G.~Hur}}}, \bibnamefont{and}
  \bibinfo{author}{\bibnamefont{{T.~S.~Monteiro}}},
  \bibinfo{journal}{Phys.~Rev.~Lett.} \textbf{\bibinfo{volume}{93}},
  \bibinfo{pages}{223002} (\bibinfo{year}{2004}).

\bibitem[{\citenamefont{{G.~Casati} et~al.}(1979)\citenamefont{{G.~Casati},
  {B.~V.~Chirikov}, {F.~M.~Izraelev}, and {J.~Ford}}}]{Casati1979}
\bibinfo{author}{\bibnamefont{{G.~Casati}}},
  \bibinfo{author}{\bibnamefont{{B.~V.~Chirikov}}},
  \bibinfo{author}{\bibnamefont{{F.~M.~Izraelev}}}, \bibnamefont{and}
  \bibinfo{author}{\bibnamefont{{J.~Ford}}}, in
  \emph{\bibinfo{booktitle}{Stochastic Behavior in Classical and Quantum
  Hamiltonian Systems}}, edited by \bibinfo{editor}{\bibnamefont{{G.~Casati}}}
  \bibnamefont{and} \bibinfo{editor}{\bibnamefont{{J.~Ford}}}
  (\bibinfo{publisher}{Springer}, \bibinfo{address}{New York},
  \bibinfo{year}{1979}).

\bibitem[{\citenamefont{{F.~M.~Izrailev} and
  {D.~L.~Shepelyanskii}}(1979)}]{Izrailev1979}
\bibinfo{author}{\bibnamefont{{F.~M.~Izrailev}}} \bibnamefont{and}
  \bibinfo{author}{\bibnamefont{{D.~L.~Shepelyanskii}}},
  \bibinfo{journal}{Sov.~Phys.~Dokl.} \textbf{\bibinfo{volume}{24}},
  \bibinfo{pages}{996} (\bibinfo{year}{1979}).

\bibitem[{\citenamefont{{F.~M.~Izrailev} and
  {D.~L.~Shepelyanskii}}(1980)}]{Izrailev1980}
\bibinfo{author}{\bibnamefont{{F.~M.~Izrailev}}} \bibnamefont{and}
  \bibinfo{author}{\bibnamefont{{D.~L.~Shepelyanskii}}},
  \bibinfo{journal}{Theor.~Math.~Phys.} \textbf{\bibinfo{volume}{47}},
  \bibinfo{pages}{553} (\bibinfo{year}{1980}).

\bibitem[{\citenamefont{{I.~Dana} et~al.}(1995)\citenamefont{{I.~Dana},
  {E.~Eisenberg}, and {N.~Shnerb}}}]{Dana1995}
\bibinfo{author}{\bibnamefont{{I.~Dana}}},
  \bibinfo{author}{\bibnamefont{{E.~Eisenberg}}}, \bibnamefont{and}
  \bibinfo{author}{\bibnamefont{{N.~Shnerb}}},
  \bibinfo{journal}{Phys.~Rev.~Lett.} \textbf{\bibinfo{volume}{74}},
  \bibinfo{pages}{686} (\bibinfo{year}{1995}).

\bibitem[{\citenamefont{{I.~Dana} et~al.}(1996)\citenamefont{{I.~Dana},
  {E.~Eisenberg}, and {N.~Shnerb}}}]{Dana1996}
\bibinfo{author}{\bibnamefont{{I.~Dana}}},
  \bibinfo{author}{\bibnamefont{{E.~Eisenberg}}}, \bibnamefont{and}
  \bibinfo{author}{\bibnamefont{{N.~Shnerb}}}, \bibinfo{journal}{Phys.~Rev.~E}
  \textbf{\bibinfo{volume}{54}}, \bibinfo{pages}{5948} (\bibinfo{year}{1996}).

\bibitem[{\citenamefont{{I.~Dana} and {D.~L.~Dorofeev}}(2005)}]{Dana2005}
\bibinfo{author}{\bibnamefont{{I.~Dana}}} \bibnamefont{and}
  \bibinfo{author}{\bibnamefont{{D.~L.~Dorofeev}}},
  \bibinfo{journal}{Phys.~Rev.~E} \textbf{\bibinfo{volume}{72}},
  \bibinfo{pages}{046205} (\bibinfo{year}{2005}).

\bibitem[{\citenamefont{{I.~Dana} and
  {D.~L.~Dorofeev}}(2006{\natexlab{a}})}]{Dana2006a}
\bibinfo{author}{\bibnamefont{{I.~Dana}}} \bibnamefont{and}
  \bibinfo{author}{\bibnamefont{{D.~L.~Dorofeev}}},
  \bibinfo{journal}{Phys.~Rev.~E} \textbf{\bibinfo{volume}{73}},
  \bibinfo{pages}{026206} (\bibinfo{year}{2006}{\natexlab{a}}).

\bibitem[{\citenamefont{{I.~Dana} and
  {D.~L.~Dorofeev}}(2006{\natexlab{b}})}]{Dana2006b}
\bibinfo{author}{\bibnamefont{{I.~Dana}}} \bibnamefont{and}
  \bibinfo{author}{\bibnamefont{{D.~L.~Dorofeev}}},
  \bibinfo{journal}{Phys.~Rev.~E} \textbf{\bibinfo{volume}{74}},
  \bibinfo{pages}{045201(R)} (\bibinfo{year}{2006}{\natexlab{b}}).

\bibitem[{\citenamefont{{S.~Wimberger}
  et~al.}(2003)\citenamefont{{S.~Wimberger}, {I.~Guarneri}, and
  {S.~Fishman}}}]{Wimberger2003}
\bibinfo{author}{\bibnamefont{{S.~Wimberger}}},
  \bibinfo{author}{\bibnamefont{{I.~Guarneri}}}, \bibnamefont{and}
  \bibinfo{author}{\bibnamefont{{S.~Fishman}}}, \bibinfo{journal}{Nonlinearity}
  \textbf{\bibinfo{volume}{16}}, \bibinfo{pages}{1381} (\bibinfo{year}{2003}).

\bibitem[{\citenamefont{{S.~Wimberger}
  et~al.}(2004)\citenamefont{{S.~Wimberger}, {I.~Guarneri}, and
  {S.~Fishman}}}]{Wimberger2004}
\bibinfo{author}{\bibnamefont{{S.~Wimberger}}},
  \bibinfo{author}{\bibnamefont{{I.~Guarneri}}}, \bibnamefont{and}
  \bibinfo{author}{\bibnamefont{{S.~Fishman}}},
  \bibinfo{journal}{Phys.~Rev.~Lett.} \textbf{\bibinfo{volume}{92}},
  \bibinfo{pages}{084102} (\bibinfo{year}{2004}).

\bibitem[{\citenamefont{{S.~Wimberger} and
  {M.~Sadgrove}}(2005)}]{Wimberger2005a}
\bibinfo{author}{\bibnamefont{{S.~Wimberger}}} \bibnamefont{and}
  \bibinfo{author}{\bibnamefont{{M.~Sadgrove}}}, \bibinfo{journal}{J.~Phys.~A}
  \textbf{\bibinfo{volume}{38}}, \bibinfo{pages}{10549} (\bibinfo{year}{2005}).

\bibitem[{\citenamefont{{L.~E.~Reichl}}(2004)}]{Reichl2004}
\bibinfo{author}{\bibnamefont{{L.~E.~Reichl}}}, \emph{\bibinfo{title}{The
  Transition to Chaos: Conservative Classical Systems and Quantum
  Manifestations}} (\bibinfo{publisher}{Springer}, \bibinfo{address}{New York},
  \bibinfo{year}{2004}), \bibinfo{edition}{2nd} ed.

\bibitem[{\citenamefont{{F.~Haake}}(2001)}]{Haake2001}
\bibinfo{author}{\bibnamefont{{F.~Haake}}}, \emph{\bibinfo{title}{Quantum
  Signatures of Chaos}} (\bibinfo{publisher}{Springer},
  \bibinfo{address}{Berlin}, \bibinfo{year}{2001}), \bibinfo{edition}{2nd} ed.

\bibitem[{\citenamefont{{M.~C.~Gutzwiller}}(1990)}]{Gutzwiller1990}
\bibinfo{author}{\bibnamefont{{M.~C.~Gutzwiller}}}, \emph{\bibinfo{title}{Chaos
  in Classical and Quantum Mechanics}} (\bibinfo{publisher}{Springer},
  \bibinfo{address}{New York}, \bibinfo{year}{1990}).

\bibitem[{\citenamefont{{M.~Saunders} et~al.}(2007)\citenamefont{{M.~Saunders},
  {P.~L.~Halkyard}, {K.~J.~Challis}, and {S.~A.~Gardiner}}}]{Saunders2007}
\bibinfo{author}{\bibnamefont{{M.~Saunders}}},
  \bibinfo{author}{\bibnamefont{{P.~L.~Halkyard}}},
  \bibinfo{author}{\bibnamefont{{K.~J.~Challis}}}, \bibnamefont{and}
  \bibinfo{author}{\bibnamefont{{S.~A.~Gardiner}}},
  \bibinfo{journal}{Phys.~Rev.~A} \textbf{\bibinfo{volume}{76}},
  \bibinfo{pages}{043415} (\bibinfo{year}{2007}).

\bibitem[{\citenamefont{{J.~{Hecker~Denschlag}}
  et~al.}(2002)\citenamefont{{J.~{Hecker~Denschlag}}, {J.~E.~Simsarian},
  {H.~H\"affner}, {C.~McKenzie}, {A.~Browaeys}, {D.~Cho}, {K.~Helmerson},
  {S.~L.~Rolston}, and {W.~D.~Phillips}}}]{Denschlag2002}
\bibinfo{author}{\bibnamefont{{J.~{Hecker~Denschlag}}}},
  \bibinfo{author}{\bibnamefont{{J.~E.~Simsarian}}},
  \bibinfo{author}{\bibnamefont{{H.~H\"affner}}},
  \bibinfo{author}{\bibnamefont{{C.~McKenzie}}},
  \bibinfo{author}{\bibnamefont{{A.~Browaeys}}},
  \bibinfo{author}{\bibnamefont{{D.~Cho}}},
  \bibinfo{author}{\bibnamefont{{K.~Helmerson}}},
  \bibinfo{author}{\bibnamefont{{S.~L.~Rolston}}}, \bibnamefont{and}
  \bibinfo{author}{\bibnamefont{{W.~D.~Phillips}}},
  \bibinfo{journal}{J.~Phys.~B} \textbf{\bibinfo{volume}{35}},
  \bibinfo{pages}{3095} (\bibinfo{year}{2002}).

\bibitem[{\citenamefont{{M.~Gustavsson} et~al.}()\citenamefont{{M.~Gustavsson},
  {E.~Haller}, {M.~J.~Mark}, {J.~G.~Danzl}, {G.~Rojas-Kopeinig}, and
  {H.-C.~N\"{a}gerl}}}]{Gustavsson2007}
\bibinfo{author}{\bibnamefont{{M.~Gustavsson}}},
  \bibinfo{author}{\bibnamefont{{E.~Haller}}},
  \bibinfo{author}{\bibnamefont{{M.~J.~Mark}}},
  \bibinfo{author}{\bibnamefont{{J.~G.~Danzl}}},
  \bibinfo{author}{\bibnamefont{{G.~Rojas-Kopeinig}}}, \bibnamefont{and}
  \bibinfo{author}{\bibnamefont{{H.-C.~N\"{a}gerl}}},
  \eprint{arXiv:0710.5083v1}.

\bibitem[{\citenamefont{{S.~Inouye} et~al.}(1998)\citenamefont{{S.~Inouye},
  {M.~R.~Andrews}, {J.~Stenger}, {H.-J.~Miesner}, {D.~M.~Stamper-Kurn}, and
  {W.~Ketterle}}}]{Inouye1998}
\bibinfo{author}{\bibnamefont{{S.~Inouye}}},
  \bibinfo{author}{\bibnamefont{{M.~R.~Andrews}}},
  \bibinfo{author}{\bibnamefont{{J.~Stenger}}},
  \bibinfo{author}{\bibnamefont{{H.-J.~Miesner}}},
  \bibinfo{author}{\bibnamefont{{D.~M.~Stamper-Kurn}}}, \bibnamefont{and}
  \bibinfo{author}{\bibnamefont{{W.~Ketterle}}}, \bibinfo{journal}{Nature
  (London)} \textbf{\bibinfo{volume}{392}}, \bibinfo{pages}{151}
  (\bibinfo{year}{1998}).

\bibitem[{\citenamefont{{J.~L.~Roberts}
  et~al.}(1998)\citenamefont{{J.~L.~Roberts}, {N.~R.~Claussen}, {J.~P.~Burke},
  {C.~H.~Greene}, {E.~A.~Cornell}, and {C.~E.~Wieman}}}]{Roberts1998}
\bibinfo{author}{\bibnamefont{{J.~L.~Roberts}}},
  \bibinfo{author}{\bibnamefont{{N.~R.~Claussen}}},
  \bibinfo{author}{\bibnamefont{{J.~P.~Burke}}},
  \bibinfo{author}{\bibnamefont{{C.~H.~Greene}}},
  \bibinfo{author}{\bibnamefont{{E.~A.~Cornell}}}, \bibnamefont{and}
  \bibinfo{author}{\bibnamefont{{C.~E.~Wieman}}},
  \bibinfo{journal}{Phys.~Rev.~Lett.} \textbf{\bibinfo{volume}{81}},
  \bibinfo{pages}{5109} (\bibinfo{year}{1998}).

\bibitem[{\citenamefont{{T.~K\"{o}hler}
  et~al.}(2006)\citenamefont{{T.~K\"{o}hler}, {K.~G\'oral}, and
  {P.~S.~Julienne}}}]{Kohler2006}
\bibinfo{author}{\bibnamefont{{T.~K\"{o}hler}}},
  \bibinfo{author}{\bibnamefont{{K.~G\'oral}}}, \bibnamefont{and}
  \bibinfo{author}{\bibnamefont{{P.~S.~Julienne}}}, \bibinfo{journal}{Rev. Mod.
  Phys.} \textbf{\bibinfo{volume}{78}}, \bibinfo{pages}{1311}
  (\bibinfo{year}{2006}).

\bibitem[{\citenamefont{{T.~Kraemer} et~al.}(2004)\citenamefont{{T.~Kraemer},
  {J.~Herbig}, {M.~Mark}, {T.~Weber}, {C.~Chin}, {H.-C.~N\"agerl}, and
  {R.~Grimm}}}]{Kraemer2004}
\bibinfo{author}{\bibnamefont{{T.~Kraemer}}},
  \bibinfo{author}{\bibnamefont{{J.~Herbig}}},
  \bibinfo{author}{\bibnamefont{{M.~Mark}}},
  \bibinfo{author}{\bibnamefont{{T.~Weber}}},
  \bibinfo{author}{\bibnamefont{{C.~Chin}}},
  \bibinfo{author}{\bibnamefont{{H.-C.~N\"agerl}}}, \bibnamefont{and}
  \bibinfo{author}{\bibnamefont{{R.~Grimm}}}, \bibinfo{journal}{Appl.~Phys.~B}
  \textbf{\bibinfo{volume}{79}}, \bibinfo{pages}{1013} (\bibinfo{year}{2004}).

\bibitem[{\citenamefont{{P.~Meystre} and {M.~Sargent~III}}(1991)}]{Meystre1991}
\bibinfo{author}{\bibnamefont{{P.~Meystre}}} \bibnamefont{and}
  \bibinfo{author}{\bibnamefont{{M.~Sargent~III}}},
  \emph{\bibinfo{title}{Elements of quantum optics}}
  (\bibinfo{publisher}{Springer-Verlag}, \bibinfo{address}{Berlin},
  \bibinfo{year}{1991}), \bibinfo{edition}{3rd} ed.

\bibitem[{\citenamefont{{C.~Kittel}}(1996)}]{Kittel1996}
\bibinfo{author}{\bibnamefont{{C.~Kittel}}}, \emph{\bibinfo{title}{Introduction
  to solid state physics}} (\bibinfo{publisher}{John Wiley \& Sons},
  \bibinfo{address}{New York}, \bibinfo{year}{1996}).

\bibitem[{\citenamefont{{R.~Bach} et~al.}(2005)\citenamefont{{R.~Bach},
  {K.~Burnett}, {M.~B.~d'Arcy}, and {S.~A.~Gardiner}}}]{Bach2005}
\bibinfo{author}{\bibnamefont{{R.~Bach}}},
  \bibinfo{author}{\bibnamefont{{K.~Burnett}}},
  \bibinfo{author}{\bibnamefont{{M.~B.~d'Arcy}}}, \bibnamefont{and}
  \bibinfo{author}{\bibnamefont{{S.~A.~Gardiner}}},
  \bibinfo{journal}{Phys.~Rev.~A} \textbf{\bibinfo{volume}{71}},
  \bibinfo{pages}{033417} (\bibinfo{year}{2005}).

\bibitem[{\citenamefont{{S.~Fishman} et~al.}(2002)\citenamefont{{S.~Fishman},
  {I.~Guarneri}, and {L.~Rebuzzini}}}]{Fishman2002}
\bibinfo{author}{\bibnamefont{{S.~Fishman}}},
  \bibinfo{author}{\bibnamefont{{I.~Guarneri}}}, \bibnamefont{and}
  \bibinfo{author}{\bibnamefont{{L.~Rebuzzini}}},
  \bibinfo{journal}{Phys.~Rev.~Lett.} \textbf{\bibinfo{volume}{89}},
  \bibinfo{pages}{084101} (\bibinfo{year}{2002}).

\bibitem[{\citenamefont{{S.~Fishman} et~al.}(2003)\citenamefont{{S.~Fishman},
  {I.~Guarneri}, and {L.~Rebuzzini}}}]{Fishman2003}
\bibinfo{author}{\bibnamefont{{S.~Fishman}}},
  \bibinfo{author}{\bibnamefont{{I.~Guarneri}}}, \bibnamefont{and}
  \bibinfo{author}{\bibnamefont{{L.~Rebuzzini}}},
  \bibinfo{journal}{J.~Stat.~Phys.} \textbf{\bibinfo{volume}{110}},
  \bibinfo{pages}{911} (\bibinfo{year}{2003}).

\bibitem[{\citenamefont{{H.~G.~Schuster}}(1995)}]{Schuster1995}
\bibinfo{author}{\bibnamefont{{H.~G.~Schuster}}},
  \emph{\bibinfo{title}{Deterministic Chaos: An Introduction}}
  (\bibinfo{publisher}{VCH}, \bibinfo{address}{Weinheim},
  \bibinfo{year}{1995}).

\bibitem[{\citenamefont{{L.~Deng} et~al.}(1999)\citenamefont{{L.~Deng},
  {E.~W.~Hagley}, {J.~Denschlag}, {J.~E.~Simsarian}, {M.~Edwards},
  {C.~W.~Clark}, {K.~Helmerson}, {S.~L.~Rolston}, and
  {W.~D.~Phillips}}}]{Deng1999}
\bibinfo{author}{\bibnamefont{{L.~Deng}}},
  \bibinfo{author}{\bibnamefont{{E.~W.~Hagley}}},
  \bibinfo{author}{\bibnamefont{{J.~Denschlag}}},
  \bibinfo{author}{\bibnamefont{{J.~E.~Simsarian}}},
  \bibinfo{author}{\bibnamefont{{M.~Edwards}}},
  \bibinfo{author}{\bibnamefont{{C.~W.~Clark}}},
  \bibinfo{author}{\bibnamefont{{K.~Helmerson}}},
  \bibinfo{author}{\bibnamefont{{S.~L.~Rolston}}}, \bibnamefont{and}
  \bibinfo{author}{\bibnamefont{{W.~D.~Phillips}}},
  \bibinfo{journal}{Phys.~Rev.~Lett.} \textbf{\bibinfo{volume}{83}},
  \bibinfo{pages}{5407} (\bibinfo{year}{1999}).

\bibitem[{\citenamefont{{P.~L.~Halkyard}
  et~al.}()\citenamefont{{P.~L.~Halkyard}, {M.~Saunders}, {S.~A.~Gardiner}, and
  {K.~J.~Challis}}}]{Halkyard2008}
\bibinfo{author}{\bibnamefont{{P.~L.~Halkyard}}},
  \bibinfo{author}{\bibnamefont{{M.~Saunders}}},
  \bibinfo{author}{\bibnamefont{{S.~A.~Gardiner}}}, \bibnamefont{and}
  \bibinfo{author}{\bibnamefont{{K.~J.~Challis}}}, \eprint{arXiv:0807.2587v1}.

\bibitem[{\citenamefont{{M.~Abramowitz} and
  {I.~A.~Stegun}}(1964)}]{Abramowitz1964}
\bibinfo{author}{\bibnamefont{{M.~Abramowitz}}} \bibnamefont{and}
  \bibinfo{author}{\bibnamefont{{I.~A.~Stegun}}},
  \emph{\bibinfo{title}{Handbook of Mathematical Functions with Formulas,
  Graphs, and Mathematical Tables}} (\bibinfo{publisher}{U.~S. Government
  Printing Office}, \bibinfo{address}{Washington}, \bibinfo{year}{1964}).

\bibitem[{\citenamefont{{C.~W.~Gardiner}}(2004)}]{Gardiner2004}
\bibinfo{author}{\bibnamefont{{C.~W.~Gardiner}}},
  \emph{\bibinfo{title}{Handbook of Stochastic Methods}}
  (\bibinfo{publisher}{Springer}, \bibinfo{address}{Berlin},
  \bibinfo{year}{2004}), \bibinfo{edition}{3rd} ed.

\bibitem[{\citenamefont{{J.~Fricke}}(1996)}]{Fricke1996}
\bibinfo{author}{\bibnamefont{{J.~Fricke}}},
  \bibinfo{journal}{Ann.~Phys.~(N.Y.)} \textbf{\bibinfo{volume}{252}},
  \bibinfo{pages}{479} (\bibinfo{year}{1996}).

\bibitem[{\citenamefont{{T.~K\"{o}hler} and {K.~Burnett}}(2002)}]{Kohler2002}
\bibinfo{author}{\bibnamefont{{T.~K\"{o}hler}}} \bibnamefont{and}
  \bibinfo{author}{\bibnamefont{{K.~Burnett}}}, \bibinfo{journal}{Phys.~Rev.~A}
  \textbf{\bibinfo{volume}{65}}, \bibinfo{pages}{033601}
  (\bibinfo{year}{2002}).

\bibitem[{\citenamefont{{A.~Peters} et~al.}(1999)\citenamefont{{A.~Peters},
  {K.~Y.~Chung}, and {S.~Chu}}}]{Peters1999}
\bibinfo{author}{\bibnamefont{{A.~Peters}}},
  \bibinfo{author}{\bibnamefont{{K.~Y.~Chung}}}, \bibnamefont{and}
  \bibinfo{author}{\bibnamefont{{S.~Chu}}}, \bibinfo{journal}{Nature~(London)}
  \textbf{\bibinfo{volume}{400}}, \bibinfo{pages}{849} (\bibinfo{year}{1999}).

\bibitem[{\citenamefont{{I.~Carusotto}
  et~al.}(2005)\citenamefont{{I.~Carusotto}, {L.~Pitaevskii}, {S.~Stringari},
  {G.~Modugno}, and {M.~Inguscio}}}]{Carusotto2005}
\bibinfo{author}{\bibnamefont{{I.~Carusotto}}},
  \bibinfo{author}{\bibnamefont{{L.~Pitaevskii}}},
  \bibinfo{author}{\bibnamefont{{S.~Stringari}}},
  \bibinfo{author}{\bibnamefont{{G.~Modugno}}}, \bibnamefont{and}
  \bibinfo{author}{\bibnamefont{{M.~Inguscio}}},
  \bibinfo{journal}{Phys.~Rev.~Lett.} \textbf{\bibinfo{volume}{95}},
  \bibinfo{pages}{093202} (\bibinfo{year}{2005}).

\bibitem[{\citenamefont{{W.~H.~Louisell}}(1973)}]{Louisell1973}
\bibinfo{author}{\bibnamefont{{W.~H.~Louisell}}}, \emph{\bibinfo{title}{Quantum
  statistical properties of radiation}} (\bibinfo{publisher}{John Wiley \&
  Sons}, \bibinfo{address}{New York}, \bibinfo{year}{1973}).

\bibitem[{\citenamefont{{T.~M.~Apostol}}(1976)}]{Apostol1976}
\bibinfo{author}{\bibnamefont{{T.~M.~Apostol}}},
  \emph{\bibinfo{title}{Introduction to Analytic Number Theory}}
  (\bibinfo{publisher}{Springer-Verlag}, \bibinfo{address}{New York},
  \bibinfo{year}{1976}).

\bibitem[{\citenamefont{{V.~Armitage} and {J.~Rogers}}(2000)}]{Armitage2000}
\bibinfo{author}{\bibnamefont{{V.~Armitage}}} \bibnamefont{and}
  \bibinfo{author}{\bibnamefont{{J.~Rogers}}}, \bibinfo{journal}{J.~Phys.~A}
  \textbf{\bibinfo{volume}{33}}, \bibinfo{pages}{5593} (\bibinfo{year}{2000}).

\bibitem[{\citenamefont{{D.~Bigourd} et~al.}(2006)\citenamefont{{D.~Bigourd},
  {B.~Chatel}, {W.~P.~Schleich}, and {B.~Girard}}}]{Bigourd2008}
\bibinfo{author}{\bibnamefont{{D.~Bigourd}}},
  \bibinfo{author}{\bibnamefont{{B.~Chatel}}},
  \bibinfo{author}{\bibnamefont{{W.~P.~Schleich}}}, \bibnamefont{and}
  \bibinfo{author}{\bibnamefont{{B.~Girard}}},
  \bibinfo{journal}{Phys.~Rev.~Lett.} \textbf{\bibinfo{volume}{100}},
  \bibinfo{pages}{030202} (\bibinfo{year}{2008}).

\bibitem[{\citenamefont{{M.~Gilowski} et~al.}(2008)\citenamefont{{M.~Gilowski},
  {T.~Wendrich}, {T.~M\"{u}ller}, {C.~Jentsch}, {W.~Ertmer}, {E.~M.~Rasel}, and
  {W.~P.~Schleich}}}]{Gilowski2008}
\bibinfo{author}{\bibnamefont{{M.~Gilowski}}},
  \bibinfo{author}{\bibnamefont{{T.~Wendrich}}},
  \bibinfo{author}{\bibnamefont{{T.~M\"{u}ller}}},
  \bibinfo{author}{\bibnamefont{{C.~Jentsch}}},
  \bibinfo{author}{\bibnamefont{{W.~Ertmer}}},
  \bibinfo{author}{\bibnamefont{{E.~M.~Rasel}}}, \bibnamefont{and}
  \bibinfo{author}{\bibnamefont{{W.~P.~Schleich}}},
  \bibinfo{journal}{Phys.~Rev.~Lett.} \textbf{\bibinfo{volume}{100}},
  \bibinfo{pages}{030201} (\bibinfo{year}{2008}).

\end{thebibliography}
\end{document}